\documentclass[prx,twocolumn,secnumarabic,balancelastpage,amsmath,amssymb,noeprint,superscriptaddress]{revtex4-2}

\usepackage{graphicx}
\usepackage[dvipsnames]{xcolor}
\usepackage[colorlinks,linkcolor=black,citecolor=black,urlcolor=black,filecolor=black]{hyperref}
\usepackage{amsmath, amssymb, amsthm}
\usepackage{wasysym}
\usepackage{lipsum}
\usepackage{outlines}
\usepackage{enumerate}
\usepackage{setspace}
\usepackage{tabularx} 
\usepackage{xcolor}
\usepackage[export]{adjustbox}
\usepackage{tikz}
\usepackage{todonotes}
\usepackage{verbatim}
\usepackage{algorithm}
\usepackage{algpseudocode}
\usepackage{verbatim}
\usepackage{wrapfig}
\usepackage{dsfont}
\usepackage{braket}
\usepackage{quantikz}
\usetikzlibrary{quantikz2}

\usepackage[normalem]{ulem}

\usetikzlibrary{decorations.pathreplacing}
\usetikzlibrary{arrows,arrows.meta,calc,shapes.geometric,shapes.misc}
\tikzset{
	>=stealth',
	help lines/.style={dashed, thick},
	important line/.style={thick},
	connection/.style={thick, dotted},
}
\DeclareMathAlphabet{\mymathbb}{U}{BOONDOX-ds}{m}{n}
\hypersetup{
    pdfstartview={FitH}, 
    colorlinks=true, 
    linkcolor=blue, 
    citecolor=blue, 
    filecolor=magenta,
    urlcolor=blue
}

\newcommand{\figref}[1]{Fig.~\ref{#1}}
\newcommand{\mC}[1]{#1}

\theoremstyle{definition}

\newif\ifptitle
\newif\ifpnumber
\newcounter{para}
\newcommand\ptitle[1]{\par\refstepcounter{para}
{\ifpnumber{\noindent\textcolor{lightgray}{\textbf{\thepara}}\indent}\fi}
{\ifptitle{\textbf{[{#1}]}}\fi}}

\begin{document}

\title{Designing Fault-Tolerant Blind Quantum Computation}

\author{Gefen~Baranes}
\thanks{These authors contributed equally}
\affiliation{Department~of~Physics~and~Research~Laboratory~of~Electronics,~Massachusetts~Institute~of~Technology,~Cambridge,~MA,~USA} 
\affiliation{Department~of~Physics,~Harvard~University,~Cambridge,~MA~02138,~USA}

\author{Iria W. Wang}
\thanks{These authors contributed equally}
\affiliation{Department~of~Physics,~Harvard~University,~Cambridge,~MA~02138,~USA}

\author{Francisco Machado}
\thanks{These authors contributed equally}
\affiliation{Department~of~Physics,~Harvard~University,~Cambridge,~MA~02138,~USA}
\affiliation{ITAMP,~Harvard-Smithsonian~Center~for~Astrophysics,~Cambridge,~MA,~USA}

\author{Aziza Suleymanzade}
\affiliation{Department~of~Physics,~Harvard~University,~Cambridge,~MA~02138,~USA}

\author{Pieter-Jan Stas}
\affiliation{Department~of~Physics,~Harvard~University,~Cambridge,~MA~02138,~USA}

\author{Yan-Cheng Wei}
\affiliation{Department~of~Physics,~Harvard~University,~Cambridge,~MA~02138,~USA}

\author{Susanne F. Yelin}
\affiliation{Department~of~Physics,~Harvard~University,~Cambridge,~MA~02138,~USA}

\author{Johannes Borregaard}
\affiliation{Department~of~Physics,~Harvard~University,~Cambridge,~MA~02138,~USA}
\affiliation{Lightsynq~Technologies~Inc.,~Brighton,~Massachusetts~02135,~USA}

\author{Mikhail~D.~Lukin}
\affiliation{Department~of~Physics,~Harvard~University,~Cambridge,~MA~02138,~USA}

\date{\today}

\begin{abstract}
    Blind quantum computing (BQC) is a computational paradigm that allows a client with limited quantum capabilities to delegate quantum computations to a more powerful server while keeping both the algorithm and data hidden. 
    However, in practice, existing BQC protocols face significant challenges when scaling to large-scale computations due to photon losses, low efficiencies, and high overheads associated with fault-tolerant operations, requiring the client to compile both logical operations and error correction primitives. 
    We use a recently demonstrated hybrid light-matter approach~[PRL 132, 150604 (2024); Science 388, 509-513 (2025)] to develop an architecture for scalable fault-tolerant blind quantum computation.
    By combining high-fidelity local gates on the server's matter qubits with delegated blind rotations using photons, we construct loss-tolerant delegated gates that enable efficient algorithm compilation strategies 
    and a scalable approach for fault-tolerant blind logical algorithms. 
    Our approach improves the error-correction threshold and increases the speed and depth of blind logical circuits. 
    Finally, we outline how this architecture can be implemented on state-of-the-art quantum hardware, including neutral atom arrays and solid-state spin defects.
    These new capabilities open up new opportunities for deep circuit blind quantum computing.
\end{abstract}

\maketitle

\section{Introduction}
\ptitle{Quantum Mechanics as a tool for more fundamental notions of security}

Quantum mechanics enables fundamentally new ways to privately communicate and compute.  
By leveraging the intrinsic randomness of quantum measurements, security of communication can, under certain conditions, be guaranteed by the laws of quantum mechanics~\cite{bennett:2014, ekert:1991}. 
This fact underpins a variety of quantum communication and quantum key distribution protocols that exhibit unconditional security guarantees~\cite{hillery:1999,barnum:2002,renner:2008,scarani:2009}.
Such security guarantees \mC{can also be} important when performing computational tasks, \mC{enabling new opportunities for applications that require secure computation~\cite{steane:1998,nielsen:2010}}.

\begin{figure}[!h]
 \centering
  \includegraphics[width=\columnwidth]{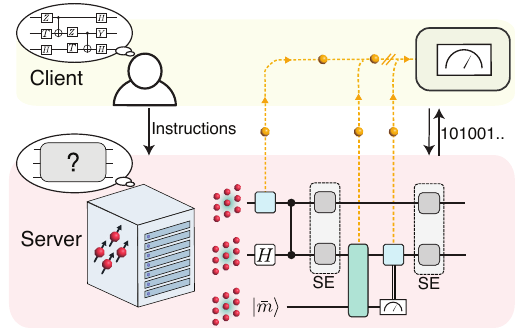} 
  \caption{
  Schematic of a blind logical computation using a hybrid light-matter platform. 
  The server has access to a matter-based quantum computer capable of performing fault-tolerant quantum computation, while the client only has access to a photonic measurement device and a classical computer. 
  The server interacts with the client via a classical channel (for instructions) and a quantum channel (for transmitting single photons).
  Through measurements on the transmitted photons, the client applies deterministic blind logical operations on the server's qubits, either via transversal gates, or gate teleportation.
  By directly manipulating the matter qubits, the server can, \emph{without compromising blindness}, complement the client's operations with non-blind logical gates, and perform the syndrome extraction (SE) required for error correction, communicating the outcomes to the client for decoding. 
  }
  \label{fig:concept} 
\end{figure}

\ptitle{Challenges }
One class of such computation protocols is known as blind quantum computing (BQC). 
Here, a client with limited quantum resources is able to securely delegate large computational tasks to a more capable quantum server without revealing details of the input, the algorithm, or the output~\cite{fitzsimons:2017}. 
In the original BQC proposal, the client was endowed with a large quantum memory, yet was only able to perform Pauli operations~\cite{childs:2005}.
Subsequent \mC{improvement} reduced the client's requirements and provided new capabilities such as verifying the correctness of the underlying computation~\cite{broadbent:2009,morimae:2012,cojocaru:2019,mahadev:2018, lee:2025,poshtvan:2025}.
These theoretical advances have been instrumental for the realization of proof-of-principle demonstrations of BQC in several different experimental platforms~\cite{barz:2012, greganti:2016, polacchi:2023, polacchi:2024,drmota:2024,wei:2024}.
Moreover, the ability to implement an arbitrary quantum computation offers a path towards fault-tolerant blind quantum computation, provided that the client compiles and implements both the logical algorithm, as well as the underlying error correction~\cite{zhao:2023, zhao:2023a, takeuchi:2017, morimae:2012}.
However, translating existing BQC approaches into practical large-scale quantum computation has remained challenging owing to the limited efficiency of quantum operations involving photonic qubits, and the large overhead associated with error correction, which further increases the overall complexity of the protocol and the associated experimental requirements.

In this article, we address these challenges by outlining a 
method for Fault-Tolerant Blind Quantum Computation (FT-BQC) within a hybrid light-matter quantum platform~[Fig.~\ref{fig:concept}].
The key idea is that 
matter-based qubits enable the server to store and fault-tolerantly manipulate the quantum state, while the photonic qubits enable the client to securely perform delegated operations.
We demonstrate how these elements enable the efficient implementation of an arbitrary blind quantum algorithm.
More importantly, this approach enables the offloading of the error correcting overhead to the server, and opens up new opportunities for the hardware-efficient design of blind quantum algorithms.

We start by introducing the hybrid light-matter architecture, which leverages a loss-tolerant photonic communication protocol for the robust and efficient implementation of blind gates. 
This architecture provides the client with the flexibility to implement both local gates and delegated operations, while preserving the blindness and verifiability of the algorithm~(Section~\ref{sec:hybrid_rules_of_the_game}).
We then explore how this approach opens up new design opportunities for the BQC algorithms.
On the one hand, it enables the client to balance between her concerns of security and efficiency.
On the other hand, it provides new opportunities for greater computational depth, expanding the space of accessible operations with fewer photonic resources (Section~\ref{sec:advantages_of_flexibility_in_circuit_design}).

Building upon the aforementioned capabilities, we then present an efficient fault-tolerant implementation of BQC. 
By deferring non-algorithmic operations to the server (including syndrome extraction and magic state preparation), our FT-BQC implementation offers a significant reduction in communication overhead.
Crucially, this improved efficiency translates into significant gains in the achievable logical circuit depths, relaxing the platform requirements for delegated fault-tolerant computation.
Using circuit-level simulations of logical algorithms on multiple surface codes, we demonstrate substantial improvements in the communication error thresholds, from $1\%$ to up to $10\%$~(Section~\ref{sec:blind_fault_tolerant_computation}).
Finally, we analyze how all these elements come together in two different state-of-the-art quantum platforms: neutral atom arrays with photonic links and solid-state spin defects in nanophotonic cavities (Section~\ref{sec:experimental_proposal_for_FTBQC}).
In both settings, we estimate the resources and time required for a large scale blind quantum computation, emphasizing the limiting factors in each platform.
The architectural, algorithmic, and error-correction tools developed in this work open up a new research direction at the intersection of error correction, circuit compilation, and delegated quantum security. 
These directions, as well as the key open questions, are discussed in Section~\ref{sec:conclusion}.

\section{Blind Quantum Computation with Light-Matter Hybrid Systems}
\label{sec:hybrid_rules_of_the_game}

\ptitle{the setting: client and server}
We consider a quantum computing architecture involving a server and a client, communicating using both a classical and a quantum photonic link. 
The server possesses extensive quantum resources and is able to perform fault-tolerant quantum computation on local matter-based qubits.
Its matter qubits can interface with a photonic link, enabling the emission of entangled photons; we term this a \emph{hybrid} light-matter architecture.
By contrast, the client has limited quantum capabilities---she is restricted to measuring incoming photonic qubits in an arbitrary basis---and can perform classical computation.

\ptitle{How to do delegated operations and what is blindness}
In this setting, the client can remotely perform quantum gates on the server's matter qubits by appropriately measuring incoming photons~\cite{morimae:2013}.
For each delegated operation, the server first entangles a matter qubit with a photonic qubit that is sent to the client.
By measuring the photon in the basis \(\ket{\pm}_\theta = (\ket{0} \pm e^{-i\theta}\ket{1})/\sqrt{2}\), 
the client imparts a rotation \(Z^s R_z(\theta)\) onto the server's qubit, where $Z$ is the Pauli operator, $R_z(\theta) = e^{i\theta Z/2}$, and $s = 0,1$ is the measurement outcome corresponding, respectively, to state $\ket{+}_\theta,\ket{-}_\theta$.
This operation is \emph{blind}: without knowledge of the client's measurement outcome $s$, the server observes a $\theta$-independent state and, thus, cannot determine \(\theta\).
This is guaranteed by the no-signaling principle~\cite{morimae:2013}.
By contrast, the client has complete knowledge of the operation; 
this protocol implements the rotation targeted by the client $R_z(\theta)$, up to a Pauli operator $Z^s$ which composes the \emph{Pauli frame}.

This simple primitive can be complemented by local gates applied directly on the matter qubits to construct more complex unitaries.
However, to enable the client to perform a deterministic operation, these local gates are constrained to implement a Clifford transformation $C$.
Such transformation maps the Pauli frame $P$ into another Pauli frame $P' = CPC^\dag$, and this map is classically computable by the client.
As a result, throughout the computation, the state remains of the form $P U \ket{0}^{\otimes n}$, where $U$ is the unitary transformation that the client intends to implement and $P$ is the instantaneous Pauli frame, dependent on the client's measurement record and the details of the computation itself.
This Pauli frame structure enables the client to hide the details of each blind operation while ensuring that its effects on both future blind operations and measurement outcomes can be classically computable.

\vspace{2mm}
To illustrate \mC{these considerations, we} explicitly construct a blind universal gate set.
An arbitrary single qubit rotation can be obtained by performing blind rotations interspersed with Hadamard gates:
\begin{equation*}
\small 
\hspace{-3mm}\begin{quantikz}[transparent, row sep={0.8cm,between origins},column sep=0.1cm]
\qw 
& \gate{X^aZ^b} 
& \gate[style={fill=red!20}]{Z^{s_1} R_z(\theta_1)} 
& \gate{H} & \gate{Z^{s_2} R_z(\theta_2)} & \gate{H} 
& \gate{Z^{s_3} R_z(\theta_3)} &
\qw
\end{quantikz}\notag
\end{equation*}
\begin{equation*}
\small
\hspace{-3mm}\begin{quantikz}[transparent, row sep={0.8cm,between origins},column sep=0.15cm]
=\hspace{1mm}\qw 
& \gate{R_z[(-1)^a\theta_1]} 
& \gate{R_x[(-1)^{b+s_1}\theta_2]} 
& \gate{R_z[(-1)^{a+s_2}\theta_3]} 
& \gate{P'} 
& \qw 
\end{quantikz}
\end{equation*}
where $P=X^aZ^b$ is the original Pauli frame, and  $P'=X^{s_2+a}Z^{s_3+s_1+b}$ constitutes the final Pauli frame of the operation based on the outcome of the three client measurements, $s_i$. 
To deterministically perform a single qubit rotation, parametrized by the Euler angles $\alpha, \beta,\gamma$, the client must adaptively choose the rotation angles $\theta_i$ depending on the previous measurement outcomes through the Pauli Frame: $\theta_1=\alpha(-1)^a$, $\theta_2 = (-1)^{s_1+b}\beta$ and $\theta_3 = (-1)^{s_2+a} \gamma$.

We can also use a single qubit blind rotation to implement a blind entangling gate as follows:
\begin{equation*}
\begin{quantikz}[transparent, row sep={0.8cm,between origins}]
\qw & \ctrl{0} & \gate{H} & \gate[style={fill=red}]{Z^s R_z(\theta)} & \gate{H} & \ctrl{0} & \qw\\
\qw & \ctrl{-1} & \qw & \qw & \qw & \ctrl{-1} & \qw
\end{quantikz}
\end{equation*}
By choosing $\theta\in\{0,\pi\}$, the resulting gate does not induce entanglement, while for $\theta\in \{\pi/2, 3\pi/2\}$, the gate generates a maximally entangling two-qubit controlled-phase gate (up to single qubit operations).
\mC{This construction has already been utilized in a recent experimental work to build a universal blind gate set~\cite{wei:2024}.}
We emphasize that, because of the Clifford nature of the local gates applied by the server, the evolution of the Pauli frame (as well as its impact on $\theta$) can be accounted for by the client.

Although these primitives enable universal blind quantum computation, the prescribed $R_z(\theta)$ blind gate is very susceptible to photon loss.
By directly entangling the emitted photon with the computational qubit, photon loss fully decoheres the matter qubit instead of applying the desired operation.
To successfully implement a circuit containing \(N\) delegated gates requires the simultaneous successful measurement of all photons.
This results in an exponential photon number overhead; on average, the server will generate \(\eta^{-N}/(1-\eta)\) photons before the computation is successfully implemented,  with \(\eta\) being the single-photon measurement success probability.
It is then critical to consider loss-tolerant approaches for implementing the blind $R_z(\theta)$ primitive. 
Such approaches are available given access to additional matter qubits.

\subsection*{Loss-Tolerant Verifiable Blind Quantum Computation}

\begin{figure}[t]
    \centering    \includegraphics[width=0.9\columnwidth]{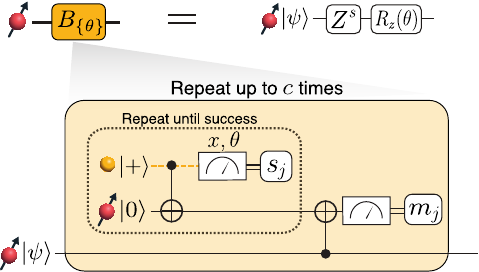}
    \caption{Loss-tolerant blind gate \( B_{\{\theta\}} \). The gate implements \( Z^s R_z(\theta) \), where \( s \in \{0,1\} \) with equal probability and \( \theta \) is chosen by the client from a set of possible angles $\{\theta\} = \{2\pi p / 2^c\}$ for integers $p = 1, ..., 2^c$. For each iteration \( j = 0,\dots,c{-}1 \), communication qubits are entangled with photons sent to the client. The server waits until the client measures a photon in the basis \( \ket{\pm}_{\phi} =  \frac{1}{\sqrt{2}}(\ket{0} \pm e^{-i\phi}\ket{1}) \) with \( \phi = 2^j\theta \), obtaining outcome \( s_j \). This teleports \( Z^{s_j}R_z(\theta) \) to the corresponding communication qubit. After entangling this communication qubit with the computation qubit and measuring it with outcome \( m_j \), the gate \( Z^{s_j}R_z[(-1)^{m_j}\theta] \) is teleported to the computation qubit. The protocol terminates if \( m_j = 0 \). The final gate is \( Z^s R_z(\theta) \), where \( s = \left(\sum_j s_j\right) \bmod 2 \) and the sum is over the iterations performed.
}
    \label{fig:hybrid_rules}
\end{figure}

By utilizing a matter qubit as a quantum memory, we now construct a deterministic, loss-tolerant blind rotation gate $B_{\{\theta\}}$ that will serve as the building block for the rest of our paper, Fig.~\ref{fig:hybrid_rules}.
The key insight is that, by using an auxiliary matter qubit to communicate with the client, the computation qubits never directly interface with the photonic link and the loss of a photon does not decohere the computation qubits.
Instead, the server can repeatedly attempt to send a photon to the client, and, upon success, teleport the blind rotation into the computational qubit~\cite{morimae:2013}.

Our proposal is as follows [\figref{fig:hybrid_rules}].
A photon is entangled with a communication qubit and sent to the client; 
if the client is unable to measure the photon, this process is retried.
Upon a successful photon measurement, a blind gate is teleported to the auxiliary communication qubit, which is then entangled with the computation qubit.
By measuring the communication qubit, the blind gate can be transferred to the computational qubit. 
However, depending on the measurement outcome of the communication qubit, this either rotates the qubit by $\theta$ or $-\theta$.
If the latter situation occurs, the entire protocol is repeated, with $\theta$ replaced by $2\theta$; now upon a successful communication qubit measurement on the server side, this implements the correct rotation by $\theta$, if not, the process is again repeated, doubling the phase in each attempt.
On average, this requires two successful photon measurements to implement the correct rotation for arbitrary angles, but the protocol might take arbitrarily many photons.
By restricting the blind angle rotations to a discrete set, \(\{\theta\} = \{2\pi p/2^c\}\), with integers \(p=1,\dots,2^c\), the number of rounds of this protocol required to implement the desired gate is at most \(c\), and on average $(2-2^{1-c})$. 

\ptitle{Explain consequences to the efficiencies}
Access to this deterministic blind gate \(B_{\{\theta\}}\) significantly reduces the photonic overhead.
With this memory-enhanced protocol, implementing \(N\) delegated gates requires on average \(N \times (2-2^{1-c})/\eta\) photons sent to the client, where $\eta/(2-2^{1-c})$ is the average efficiency of a $B_{\{\theta\}}$ gate. The expected number of photons thus scales linearly with the size of the computation, yielding an exponential improvement in efficiency over a memory-less approach.

A few remarks are in order.
First, while the primitive described in this section requires two-way communication between the server and the client, blindness can still be guaranteed.
As discussed in detail in Ref.~\cite{morimae:2013}, a measurement-only protocol has the advantage that the server cannot deduce any information of the client's measurements without knowledge of the measurement result.
However, owing to the interactive nature of the loss-tolerant protocol, information about the measurement angles can, in principle, leak to the server via the classical protocol.
The client can ensure no leakage occurs, in a device-independent way, by performing a loophole-free Bell test during the implementation of a loss-tolerant blind gate~\cite{reichardt:2013}.
Alternatively, the client can certify that the efficiency of her device's measurement is independent of the measurement basis and no additional information is leaked.

Second, our protocol is also compatible with previous protocols for blind verification~\cite{hayashi:2015}.
The key idea is to extend the quantum computation with a set of test qubits which are manipulated by local non-blind gates and non-local blind gates, yet evolve trivially under the entire computation.
By passing all of the checks on the test qubits, the client has high confidence that the server has performed the correct computation.

\mC{Finally, we emphasize that the elements described in this section are already available in different state-of-the-art experimental platforms for blind computation, namely, trapped ions \cite{drmota:2024} and solid state defects in nanophotonic cavities \cite{wei:2024}.
In both cases, different qubits can be used for interfacing with the photonic link and storing the quantum state for computation, enabling the server to retry the photon emission step until success.
}

\begin{figure*}[t]
    \centering    
    \includegraphics[width=2\columnwidth]{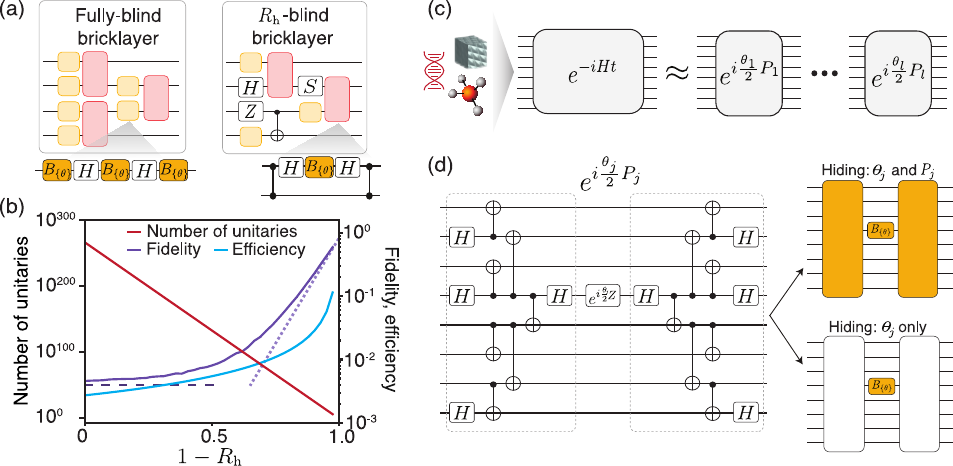}
    \caption{Practical Security.
    (a) The fully-blind bricklayer circuit constructed from universal single-qubit blind rotations (yellow) and two-qubit blind entangling gates (red). The $R_{\text{h}}$-blind bricklayer has a fraction $1-R_{\text{h}}$ of blind gates replaced with local, non-blind gates.
    (b) A trade-off between fidelity, efficiency, and security for a circuit of depth 12 and $n=8$ qubits, where $1-R_\text{h} = 0$ corresponds to a fully blind circuit. The fidelity is defined between a noisy state $\rho_\text{n}$ and an ideal noiseless state $\rho_\text{i}$ as $F(\rho_\text{n},\rho_\text{i}) = \left(\text{tr}\sqrt{\sqrt{\rho_\text{n}}\rho_\text{i}\sqrt{\rho_\text{n}}}\right)^2$. 
    The dashed line corresponds to the fully decohered limit, $F(\rho_\text{n},\rho_\text{i}) = 2^{-n})$, while the dotted line is a lower bound on the fidelity based upon the local and communication error rates~\cite{si:}. 
    The probability of a local Pauli error on a gate is set to $5\%$ for $B_{\{\theta\}}$ and $0.5\%$ for a 2-qubit local gate. 
    Efficiency is defined as the reciprocal of the expected number of photons required by a circuit, with the efficiency of a $B_{\{\theta\}}$ gate as $\eta = 80\%$. The number of unitaries is $|\{\theta\}|^N$, for a circuit of $N$ blind $B_{\{\theta\}}$ gates.
    (c) An illustration of Hamiltonian simulation approximated as the Trotter decomposition and using Pauli strings.
    (d) An example of how an $n$-qubit Pauli rotation is implemented, where $P_j = XZXZXXXZ$. 
    This may be implemented blindly by either hiding both $\theta_j$ and $P_j$ (top), which requires $\text{poly}(n)$ blind gates, or hiding only $\theta_j$, which requires one blind gate (bottom).
    }
    \label{fig:Fig_circ_design_sec}
\end{figure*}

\section{Hybrid Circuit Design}
\label{sec:advantages_of_flexibility_in_circuit_design}

The delegated blind and local gate elements introduced in the previous section can be combined in various ways, enabling highly flexible circuit designs.
In this section, we explore how these tools enable the client to design circuits that optimize resource efficiency while enabling a practical level of security and providing access to a diverse set of unitaries.

\subsection*{Practical Circuit-Level Security}
\label{sec:practical_security}

In Section~\ref{sec:hybrid_rules_of_the_game}, we describe how the rotation angles of individual blind gates are completely hidden from the server; this property is referred to as \emph{blindness}. 
We now explore how this property translates to the blindness and security of an entire circuit.
Previous works have considered circuits composed exclusively of blind gates, often focusing on bricklayer circuits as concrete examples~\cite{broadbent:2009,fitzsimons:2017}. While the individual gates in these circuits are blind, the overall circuit structure is exposed, thereby revealing the set of implementable unitaries.   
In the context of the matter-based approach, the use of blind and local gates offers greater flexibility in circuit design; however, the client must carefully consider what information is revealed by a particular circuit design. 

The simplest way to characterize the information revealed to the server is to consider the set of unitaries that a blind circuit could potentially implement. 
With the matter-based architecture, the client may choose to replace some blind operations with local gates, known to the server,  in order to reduce the communication overhead required for delegated computation, resulting in greater efficiency and fidelity for the quantum computation.
However, this trade-off comes at a cost: reducing the number of blind operations further constrains the set of implementable unitaries, thereby weakening the client's \emph{security}. In particular, the server gains more information and becomes better positioned to infer the client’s target unitary.

This trade-off is made explicit by analyzing how replacing blind gates with local, non-blind gates affects fidelity, efficiency, and security. 
We start with a fully-blind bricklayer circuit consisting of layers of single-qubit universal rotations and two-qubit entangling gates~[Fig.~\ref{fig:Fig_circ_design_sec}(a)].
Then, we replace a fraction \(1 - R_\text{h}\) of the blind gates in the circuit with local gates known to the server, where \(R_\text{h}\) denotes the \emph{hiding fraction}—the proportion of gates in the circuit that remain blind.
Reducing the hiding fraction \(R_\text{h}\) improves both the fidelity and efficiency of the computation, but at the cost of reducing the number of implementable unitaries~[Fig.~\ref{fig:Fig_circ_design_sec}(b)].
More specifically, for a computation using a total of $N_\text{tot}$ blind and non-blind gates, the efficiency of the computation improves as $\sim \frac{\eta}{N_\text{tot}R_{\text{h}}}$, and the number of unitaries decreases as $2^{c N_\text{tot}R_{\text{h}}}$.
The behavior of the fidelity is harder to capture analytically. It is lower bounded by $\sim (f_{\text{blind}}/f_{\text{local}})^{N_\text{tot}R_{\text{h}}}$, where $f_{\text{blind}} [f_{\text{local}}]$ is the fidelity of a blind [local] gate~\cite{si:}.
This lower bound exhibits an exponential dependence on the hiding ratio that accurately captures the fidelity scaling of our simulated circuits as $1-R_\text{h}$ approaches 1. For $1-R_\text{h}$ near 0, the fidelity asymptotically approaches $1/2^n$ for $n$ qubits, corresponding to the maximally mixed state~[Fig.~\ref{fig:Fig_circ_design_sec}(b, dashed)].

Thus, \(R_\text{h}\) serves as a tunable parameter that enables the client to navigate the trade-off between resource performance and security, adapting the circuit design to suit her specific requirements.

\vspace{5mm}

While the size of the set of implementable unitaries provides a general characterization of the information revealed to the server, it may not be sufficient in practical scenarios. 
In real-world contexts, this set may contain only a small number of ``interesting'' unitaries, i.e., unitaries that the client is plausibly interested in implementing. 
This could occur because the set of accessible quantum algorithms may be limited, or the client's identity may imply which algorithms she is likely to implement.

For example, suppose there are \( K \) unitaries that are of interest to the client.
If the server is aware of this set, then the number of unitaries the client is likely to implement is significantly smaller than the total space of possible unitaries. 
In this case, the number of plausible unitaries is at most \( K \), and is maximized when the client’s circuit design is capable of implementing all \( K \) target unitaries.
To construct such a circuit, the client must account for the depths, numbers of qubits, and structural characteristics of all \( K \) interesting unitaries.
A naive approach might be to compile the computation into a sufficiently deep, fully-blind bricklayer circuit, but this would be highly inefficient in practice.
Instead, the matter-based architecture offers a more efficient alternative by enabling the client to reveal parts of the computation in a targeted manner, balancing security with resource constraints.

To demonstrate this concept, we construct an example whereby the careful encoding of the problem into a blind circuit enables the client to completely and efficiently hide a set of unitaries.
More specifically, we focus on Hamiltonian simulation, which is a promising application of quantum computing, particularly in quantum chemistry and materials science~\cite{daley:2022, maskara:2025}. 
If the client does research in drug development, it may be obvious that the computation being performed is likely a Hamiltonian simulation via Trotterization~\cite{trotter:1959, mcardle:2020}. That is, the client approximates the continuous time evolution of a Hamiltonian $H$ by executing a sequence of \( n \)-qubit Pauli rotations of the form  
\begin{align}
    U(t) = e^{-iHt} \approx \left[\prod_j \exp\left(-i \frac{h_j t}{n_{\text{Trotter}}}  P_j \right)\right]^{n_\text{Trotter}},
\end{align}
where \( H \) is a linear combination of the Pauli strings \( P_j \),  
\begin{align}
    H = \sum_j h_j P_j,
\end{align}
and \( n_\text{Trotter} \in \mathbb{N}^+ \) denotes the number of Trotter steps. This process is illustrated in Fig.~\ref{fig:Fig_circ_design_sec}(c). In this scenario, all interesting unitaries correspond to Hamiltonian simulations. The client therefore may not necessarily aim to obscure the fact that Hamiltonian simulation is being performed, but rather to conceal specific details of the simulation such as the structure and parameters of the Hamiltonian.

A method for implementing an \( n \)-qubit Pauli rotation \( e^{i\frac{\theta_j}{2} P_j} \) is shown in Fig.~\ref{fig:Fig_circ_design_sec}(d)~\cite{haah:2024}, where a single-qubit rotation about the $\hat{z}$-axis is surrounded by Clifford entangling gates. The layers of Clifford gates are determined by the Pauli string \( P_j \), while the rotation angle \( \theta_j \) is encoded in the single-qubit \( \hat{z} \) rotation. In the context of Trotterization, this corresponds to \( \theta_j = -2h_j t / n_\text{Trotter} \).

In a fully blind architecture, implementing this Pauli rotation requires \( \text{poly}(n) \) blind gates where $n$ is the number of qubits. However, if the client’s objective is solely to obscure the coefficient \( h_i \), and thus the rotation angle \( \theta_i \), only the single-qubit \( \hat{z} \) rotation needs to be implemented blindly; we call this circuit the $\theta$-blind Pauli rotation. Consequently, performing an \( n \)-qubit Pauli rotation while keeping only \( \theta_i \) blind requires just a single blind gate, yielding a polynomial reduction in the required blind resources. 

Thus, when the server has some prior knowledge about the client and the computation, the client can strategically determine which aspects of the computation to conceal.
We note that in general, solving such a design problem is an important open question, and is specific to the settings under which BQC is performed. 
One approach to this challenge is presented by Ref.~\cite{poshtvan:2025} in the context of Prepare-and-Send BQC~\cite{broadbent:2009} from a more general and information-theoretic approach.

\subsection*{Efficient and Expressible Circuits}
\label{sec:efficient_and_expressible_circuits}
Access to local and non-local gates also enables the client to more efficiently implement a varied set of unitaries using fewer delegated blind gates.

To quantify this, we introduce the notion of \emph{expressibility}, which characterizes the ability of a circuit to generate a diverse and representative set of unitaries across the space of all possible \( 2^n \times 2^n \) unitaries~\cite{sim:2019,nakaji:2021,tangpanitanon:2020}. 
More specifically, each blind circuit defines an ensemble of unitaries, which is generated by the set of all possible measurement angles that may be chosen by the client.
A circuit with maximum expressibility generates a uniform distribution over the space of all possible unitaries, known as the Haar ensemble. 
Conversely, a circuit with minimal expressibility implements only a single unitary, i.e., a circuit with no blind gates. 
Thus, high expressibility indicates that the client has access to a greater variety of unitaries.

Due to its ability to implement an \( n \)-qubit operation using only a single blind gate, the Pauli rotation circuit in Fig.~\ref{fig:Fig_circ_design_sec}(d) allows for the construction of a highly expressible circuit of the form \( e^{i \theta_1 P_1}e^{i \theta_2 P_2} \dots e^{i \theta_l P_l} \), where only the angles \( \theta_i \) remain hidden. We evaluate this numerically by benchmarking Pauli rotation circuits against fully-blind bricklayer circuits.

\begin{figure}
    \centering    \includegraphics[width=\columnwidth]{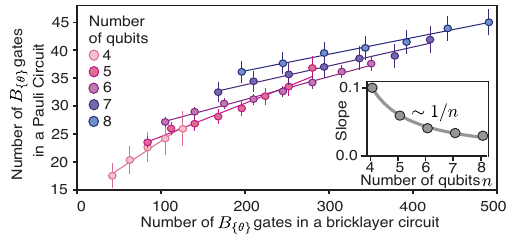}
    \caption{Expressiblity of Pauli rotations and Bricklayer circuits. Points represent the number of $B_{\{\theta\}}$ gates in a bricklayer circuit and the number of $B_{\{\theta\}}$ gates in a sequence of $\theta$-blind Pauli rotations required to achieve approximately equal expressiblities, as measured by the $k=2$ frame potential. 
    For given number of qubits these points scale linearly. The slopes scale approximately as $1/n$.}
\label{fig:Fig_circ_design_express}
\end{figure}

We quantitatively compare these two circuits using the frame potential, which is commonly used as a measure of expressiblity~\cite{hunter-jones:2019,liu:2022,nakaji:2021,sim:2019}. 
The \( k \)th \emph{frame potential} is defined as  
\begin{align}
    \mathcal{F}_\mathcal{E}^{(k)} = |\mathcal{E}|^{-2} \sum_{U,V \in \mathcal{E}} |\text{Tr}(U^\dagger V)|^{2k},
\end{align}
where \( \mathcal{E} \) represents the ensemble of all unitaries that may be generated by a given circuit, and \( k \in \mathbb{N}^+ \). A lower frame potential indicates greater expressibility, with the minimum achieved when $\mathcal{E}$ forms a $k$-design, i.e. the \( k \)-th moment of \( \mathcal{E} \) matches that of the Haar distribution~\cite{hunter-jones:2019,liu:2022}.

The frame potential is numerically estimated by sampling unitaries from the ensemble~\cite{hunter-jones:2019,liu:2022}. If the ensemble \( \mathcal{E}_\text{B} \), generated by a bricklayer circuit, and the ensemble \( \mathcal{E}_\text{P} \), generated by Pauli rotations, exhibit approximately the same frame potential for some \( k \), we conclude that they have comparable expressibility up to degree \( k \) (Appendix~\ref{app:expressibility}).

For a given \( n \) qubits, we fix the number of layers \( l_\text{B} \) in the bricklayer ansatz and determine the number of layers \( l_\text{P} \) in the Pauli rotation ansatz required to achieve the same frame potential for a given \( k \).
These layer counts are then converted into the number of \( B_{\{\theta\}} \) gates required to implement each circuit, as shown in Fig.~\ref{fig:Fig_circ_design_express}, exhibiting a linear trend whose slope decreases as \( 1/n \). Intuitively, this is because frame potentials for both families of bricklayer and $\theta$-blind Pauli rotation circuits decay exponentially with the \emph{number of layers} in a circuit.
However, the number of blind gates per layer in a bricklayer circuit scales linearly with the number of qubits, while in $\theta$-blind Pauli rotations the number of blind gates per layer is constant.
This analysis demonstrates that an equivalent degree of expressibility can be attained with a scalable reduction in the number of photons, leveraging the flexibility afforded by matter-based computing. 

Note that, while these observations were obtained upon a particular choice of blind single- and two-qubit gates, motivated $B_{\{\theta\}}$, these results are generic.
More specifically, because the enhancement in expressability is a feature of the interplay between non-local blind gates and local gates, adapting these tools to the compilation of circuits at the \emph{logical level} (as discussed in detail in the next section) would lead to similar improvements in expressibility.

\section{Fault-Tolerant Blind Quantum Computation (FT-BQC)}
\label{sec:blind_fault_tolerant_computation}

\ptitle{intro - core ideas}
We now demonstrate how our hybrid light-matter architecture enables scalable, fault-tolerant blind quantum computation. 
The central idea is to store quantum information in the server’s matter qubits using quantum error-correcting codes, with the server fully responsible for implementing the quantum error correction stack. 
The client's role is then limited to executing blind gates via photon measurements and performing decoding based on communicated syndromes. 
This clear division of roles, where matter qubits are leveraged for encoding and manipulating logical information, while photonic qubits are utilized to perform physical blind gates, significantly reduces complexity for the client while maintaining algorithmic blindness.

\begin{figure*}[t]
 \centering
  \includegraphics[width=\textwidth]{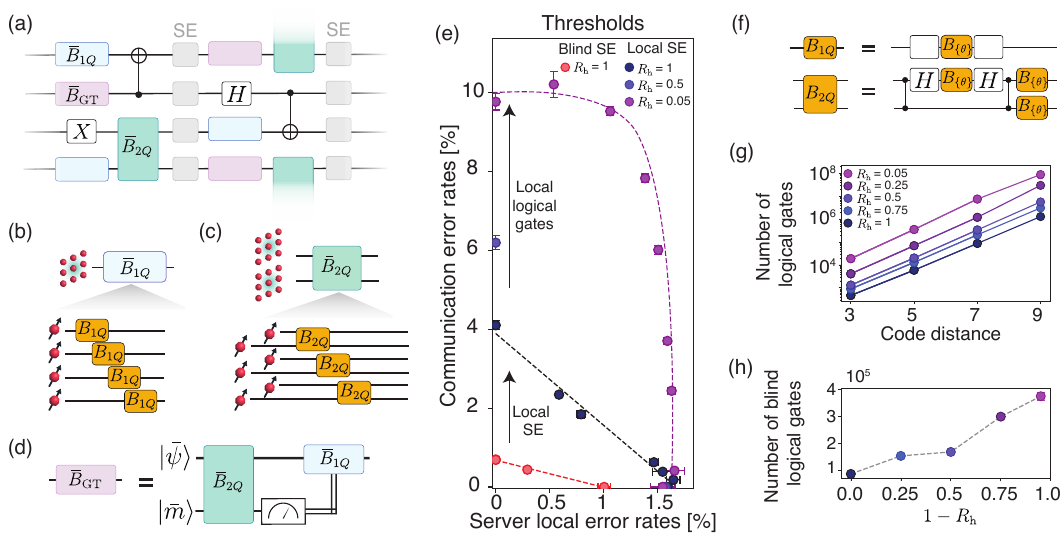}
  \caption{
  Blind fault-tolerant quantum computation.  
  (a) Deep random blind logical algorithm used for numerical simulation, consisting of 20 layers of transversal gates---a fraction \(R_{\text{h}}\) of them blind---, followed by a syndrome extraction (SE) round after each gate layer. 
  (b, c) Blind implementation of transversal blind 1-qubit gates ($\bar{B}_{1Q}$) and transversal blind 2-qubit gates ($\bar{B}_{2Q}$). Each matter qubit entangles with a photonic qubit, which is measured by the client to enact the blind transversal operation.   
  (d) Blind implementation of a single-qubit blind gate teleportation ($\bar{B}_{\text{GT}}$) utilizing the transversal blind gates in (a) and a magic state $\ket{m}$ prepared by the server. 
  (e) Phase space of correctable errors, showing thresholds for both communication errors and local server errors. Red points represent blind SE implementation, while other points correspond to local SE on the server. 
  As $R_\text{h}$ decreases, the communication error thresholds increase.
  (f) Physical implementation of blind $B_{1Q}$ and $B_{2Q}$ gates used in the numerical simulations. 
  We model the error as follows: $B_{\theta}$ gates are followed by a single-qubit communication error channel with rate $\epsilon_{\text{comm}}$, while $CZ$ gates are followed by a two-qubit server local error channel with rate $\epsilon_{\text{loc}}$.
  (g) Total number of gates executable while maintaining a final logical fidelity greater than 0.5 as a function of code distance. 
  (h) Total number of blind gates as a function of \(1 - R_{\text{h}}\) for \(d = 7\), highlighting improvements with reduced hiding.
  In (g) and (h), the error rates considered are $(\epsilon_{\text{comm}}, \epsilon_{\text{loc}}) = (0.5\%, 0\%)$.
  }
  \label{fig:QEC} 
\end{figure*}

\subsection*{Core concepts in FT-BQC}
\ptitle{Workflow: QECC selection and gate set}
Implementing fault-tolerant blind quantum computation begins by selecting an appropriate Quantum Error Correcting (QEC) code to encode quantum information into the server’s matter qubits. 
This choice can be driven by the client's computational objectives, or by the server's hardware constraints. 
Specifically, the client's desired logical circuit and the associated logical gate set may determine the optimal QEC code for efficient execution. 
Alternatively, hardware limitations—such as available native gates, qubit connectivity, or error rates—can dictate the choice of QEC codes, requiring subsequent circuit compilation tailored to the selected code. 

\ptitle{Blind implementation of transversal gates}
The choice of error correcting code determines the set of transversal gates available to the server and the client.
Such transversal logical gates can be implemented by acting simultaneously on all physical qubits of an $n$-qubit QEC code: $\bar{U}=\prod_{j=1}^n U_j$. 
This property has two important consequences.
First, they do not spread errors within a logical block, making them a cornerstone in fault‑tolerant circuit design.
Second, they are easily implementable in our architecture; each individual gate can be applied using a combination of local known Clifford and single qubit loss-tolerant blind rotation $B_{\{\theta\}}$~[Fig.~\ref{fig:QEC}(b,c)]. 
For example, the Steane code supports transversal Clifford operations (as illustrated in detail in Appendix ~\ref{app:Steane_code}), while higher dimensional color codes allow certain transversal non-Clifford gates~\cite{kubica:2015}.
Note that due to the discrete nature of logical gates, only a small set of angles $\theta$ is needed, enabling deterministic implementation using just a few successful photons per data qubit.

\ptitle{implementation of two-qubit gates}
Transversality is not limited to single-qubit gates, with certain codes also supporting transversal entangling gates across different error correcting codes.
Two-qubit transversal gates have been experimentally realized on neutral atom and trapped ion platforms~\cite{ryan-anderson:2024, bluvstein:2024}, enabling constant-time logical operations~\cite{zhou:2024,cain:2024}.
As in the single qubit transversal gate case, two-qubit gates can be made blind by constructing the relevant blind physical gates $U_i$ that are applied then to the different qubit of the error correcting code.

\ptitle{non-transversal gates}
Since transversal gates alone are insufficient for universal computation~\cite{eastin:2009}, we complement our transversal operations with a blind magic-state teleportation gadget, as illustrated in Fig.~\ref{fig:QEC}(d). 
The server prepares and distills magic states, which can then be teleported into the computational qubits via blind single- and two-qubit operations~[Fig.~\ref{fig:QEC}(b,c)].
The specifics of the magic state teleportation circuit depend on the set of gates available in the QEC code; 
for example, in the Steane code, one can leverage transversal $CX$ and $S$ gates to teleport a $T$ gate using only two delegated transversal operations.
Crucially, this blind teleportation scheme decouples the preparation of the magic state (which can be performed without communication to the client), and the teleportation of the gate itself.

Moreover, although not the focus of our subsequent analysis, we note that our architecture is also compatible with other methods for applying multi-qubit gates, most notably lattice surgery~\cite{horsman:2012}. 
In this case, a single logical entangling operation involves $O(d)$ rounds of blind entangling gates applied to seam qubits, where $d$ is the code distance.

\ptitle{local gates improvement}
Access to local gates on the server 
directly unlocks two important simplifications in our protocol.
First, it enables the client to instruct the server to implement non-blind logical operations locally, increasing the flexibility and efficiency of the computation~(see Section~\ref{sec:practical_security}). 

\ptitle{Local stabilizer checks and syndrome extraction}
Second, it enables the server to directly perform syndrome extraction (SE), i.e. measuring the stabilizer of the QEC code.
Once the stabilizers are measured, their outcomes are communicated to the client for decoding.
This process does not compromise blindness, as each syndrome extraction reveals no information about the logical state.
Moreover, as discussed in Section~\ref{sec:hybrid_rules_of_the_game}, no action by the server—including stabilizer measurements—can compromise blindness.

\ptitle{connecting all points and talk about num of photons saved. End by: we now proceed to quantify it.}
Considering all elements presented above, our hybrid architecture significantly reduces the number of photonic qubits required for logical blind computation. 
Local SE eliminates the need for photons when performing stabilizer checks;
since stabilizer measurements are required after each computational step, the reduction scales with both the depth and the number of physical qubits.
Similarly, local non-blind Clifford gates eliminate the need for photonic communication for many logical operations (as discussed in Section~\ref{sec:advantages_of_flexibility_in_circuit_design}). 
Finally, offloading magic state generation to the server removes the photonic cost typically associated with non-Clifford operations, which consume a significant portion of the computational resources~\cite{fowler:2012,litinski:2019}.
Offloading this process to the server improves both the speed and efficiency of error correction. 
Unlike previous proposals where the client must compile and execute the full fault-tolerant circuit~\cite{morimae:2012, takeuchi:2017}, our approach enables the client to only be involved when implementing blind logical operations.

\ptitle{conclusion - quantify}
Consequently, our approach requires substantially fewer photons and leverages higher fidelity local matter-based gates compared to photon-based operations. 
Reducing photonic overhead directly translates into fewer error-prone operations, which enables improved fault-tolerance thresholds.

\subsection*{Improved error correcting threshold for FT-BQC}

\ptitle{simulations introductory details}
We now quantify how improvements in efficiency translate into benefits of the underlying error correcting thresholds.
We perform extensive circuit-level QEC simulations of logical random quantum circuits incorporating transversal gates followed by SE, accessing the fidelity of the circuit as a function of both local and non-local errors [Fig.~\ref{fig:QEC}(a)].
In our simulations, utilizing the surface code, local operations on the server are subject to two-qubit depolarizing noise with rate $\epsilon_{\text{loc}}$, while client-delegated blind rotations experience single-qubit depolarizing errors with rate $\epsilon_{\text{comm}}$. 
Decoding is conducted using a most-likely-error correlated decoder, which captures correlated errors across logical qubits~\cite{cain:2024,baranes:2025}. 
By systematically analyzing how logical error rates scale with physical error probabilities and code distances, we extract error thresholds for our architecture (see  Appendix~\ref{app:QEC_simulations} for details).

\ptitle{Compare local SE vs blind SE}
We begin by analyzing the scenario where SE is implemented using delegated blind gates.
This approach mirrors previous proposals for fault-tolerant blind quantum computation, whereby the client compiles the entire computation~\cite{zhao:2023, zhao:2023a, takeuchi:2017, morimae:2012}.
In this setting, the system exhibits error thresholds of around $1\%$ for both the local two-qubit gate operations, as well as the non-local communication errors~[Fig.~\ref{fig:QEC}(e), red].

Crucially, when leveraging the server's ability to locally perform SE, these thresholds improve significantly.
The error threshold for photonic non-local gates becomes $4\%$ while the local error threshold also increases to $1.5\%$, in line with a substantial reduction in the number of corresponding gates~[Fig.~\ref{fig:QEC}(e), blue].
Such results significantly relax the requirements for performing blind quantum computation below threshold, both for local operation and communication link fidelity.
At the same time, they quantify the trade-off between the fidelity of local operations and the fidelity of the photonic link, increasing the landscape of promising paths towards fault-tolerant delegated computation.

\ptitle{Compare hiding fraction results: present fraction of hiding again, refer to thresholds, explain the 2D and 3D thresholds intuition: valid for any QEC code.}
A large portion of these gains is a result of the reduction in the number of delegated photonic gates.
Building upon this insight, we can leverage the ability to perform local logical gates on the server to further improve the error correcting thresholds.
As discussed in Section~\ref{sec:advantages_of_flexibility_in_circuit_design}, partially revealing logical gates enables a strategic interplay between fidelity and resource efficiency, captured by the hiding fraction~$R_{\text{h}}$. 
We now quantify how $R_{\text{h}}$ governs the achievable error-correction thresholds and, consequently, the potential depth of a quantum algorithm. 
As shown in Fig.~\ref{fig:QEC}(e), reducing the hiding fraction from $1$ to $0.05$ substantially increases the communication-error threshold from $4\%$ to $10\%$.
Critically, it also reshapes the fault-tolerant region in the ($\epsilon_{\text{local}},\epsilon_{\text{comm}}$) plane.
At $R_\text{h}=1$, the threshold boundary exhibits a linear trade-off between local and communication error rates. 
In contrast, for $R_\text{h}=0.05$, the boundary bows significantly outwards, 
extending towards the maximum tolerable thresholds defined by each independent error source (local and communication), thus greatly expanding the parameter space for fault-tolerant operation.

This behavior can be intuitively understood by considering the structure of errors in the decoding graph. 
Errors occur on space-like edges, primarily introduced during logical gate implementations (contributing to $\epsilon_{\mathrm{comm}}$ when gates are hidden or $\epsilon_{\mathrm{local}}$ when they are known), and on time-like edges, primarily from faulty SE measurements (contributing to $\epsilon_{\mathrm{local}}$). 
Let us examine the intercepts of the threshold boundaries in Fig.~\ref{fig:QEC}(e) with the axes. 
The vertical intercept ($\epsilon_{\mathrm{local}}=0$) represents the threshold against pure communication errors. 
In this limit, the time-like edges associated with SE have zero error probability. 
While the computation unfolds in 3D spacetime, the dominant failure mechanism involves correcting the spatial spread of errors introduced by gates, making the effective decoding problem locally resemble that of a 2D system. 
The corresponding 2D surface code phenomenological threshold is approximately $\epsilon_{\mathrm{th, 2D}} \approx 10\%$~\cite{honecker:2001, ramette:2024}. 
When all gates are hidden ($R_{\text{h}}=1$), our gate construction [Fig.~\ref{fig:QEC}(f)] introduces, on average, $\sim 2.5$ potential communication error locations per qubit per layer. 
The effective threshold is therefore suppressed to $\epsilon_{\mathrm{th, 2D}} / 2.5 \approx 4\%$, consistent with our simulations. 
Conversely, when $R_{\text{h}} \ll 1$, the \textit{average} density of communication errors per physical qubit across the computation is significantly reduced, approaching $\sim 1$, allowing the threshold to approach the full $\epsilon_{\mathrm{th, 2D}} \approx 10\%$. 

Similarly, the horizontal intercept ($\epsilon_{\mathrm{comm}}=0$) corresponds to purely local SE errors. 
This is the standard scenario for surface code fault tolerance, governed by the full 3D decoding graph, with a phenomenological threshold $\epsilon_{\mathrm{th, 3D}} \approx 3\%$~\cite{honecker:2001}. 
During each SE round, errors can occur on multiple ($\sim 2$ on average) physical data or measurement qubits involved in checking a given stabilizer, effectively scaling the threshold to $\epsilon_{\mathrm{th, 3D}} / 2 \approx 1.5\%$, again matching our numerical findings. 
The threshold boundary for $\epsilon_{\mathrm{local}}, \epsilon_{\mathrm{comm}} > 0$ interpolates between these limiting cases.

We now explain the changing \textit{shape} of the threshold boundary with $R_{\text{h}}$. 
At high hiding fractions ($R_{\text{h}}=1$), the frequent execution of hidden gates introduces dense layers of potential space-like errors. 
These errors are more likely to correlate or combine with the time-like errors originating from local SE operations, creating complex error chains that can span the code and cause logical failures. 
This increased probability of forming logical errors from combinations of both error types results in a stricter, nearly linear trade-off between tolerable $\epsilon_{\mathrm{local}}$ and $\epsilon_{\mathrm{comm}}$. 
Conversely, reducing the hiding fraction ($R_{\text{h}} \ll 1$) renders the space-like communication errors sparse in the time dimension. 
These sparse errors are far less likely to align with local time-like errors to form deleterious, large-scale error chains. 
This situation is analogous to the behavior of modular architectures described in Refs.~\cite{ramette:2024, sinclair:2024}, where errors on the lower-dimensional "seams" between modules are significantly less damaging than bulk errors, even when both error rates are substantial. 
Because the sparse communication errors and the bulk local errors are less likely to correlate to cause logical failure, the system can tolerate higher rates of \textit{both} simultaneously, leading to the outward bowing of the threshold boundary observed at low~$R_{\text{h}}$.

These observed features---distinct threshold behaviors dominated by 2D or 3D characteristics depending on the error source, and the ability to tune between them---are not specific to the surface code. 
Many QEC codes, particularly CSS codes, exhibit decoding problems mappable to statistical mechanics models where error dimension and density critically influence the existence of a fault-tolerant threshold~\cite{dennis:2002}. 
The underlying principle is geometric: sparse, lower-dimensional error sources (like our communication errors at low~$R_{\text{h}}$) are inherently less likely to create logical failures compared to dense, higher-dimensional errors (like local errors occurring throughout the spacetime volume). 
Therefore, we expect the observed improvements to be a generic feature. 
High-fidelity local operations can establish a robust fault-tolerance baseline, potentially governed by a higher, 2D-like threshold regime associated with sparse non-local errors. 
The hiding fraction then acts as a control knob, tuning the density of these non-local errors and thus interpolating the overall threshold behavior between the 2D and 3D limits, allowing optimization based on hardware capabilities and algorithmic requirements.

\ptitle{Impact of hiding fraction on achievable computational size}
Finally, we leverage our simulation results and an analytical framework described in Appendix~\ref{app:QEC_simulations} to estimate the maximum achievable number of blind gates that an algorithm can support while maintaining a logical fidelity above $50\%$. 
We begin by considering the effect of code distance on the size of the achievable computation, Fig.~\ref{fig:QEC}(g).
For $(\epsilon_{\text{comm}}, \epsilon_{\text{local}}) = (0.005,0)$, increasing the code distance enables exponential enhancement in algorithmic depth, as expected when operating below the error-correction threshold. 
Interestingly, reducing the hiding fraction $R_{\text{h}}$—thereby introducing non-blind gates into the computation—increases the size of the computation not only by the addition of these non-blind gates, but, more importantly, by reducing the logical error rate of the computation, enabling significantly more blind operations.
At fixed code distances, lower hiding fractions facilitate deeper blind computations~[Fig.~\ref{fig:QEC}(h)]. 
This finding emphasizes the beneficial interplay between blind and non-blind gates, setting the stage for practical implementation on real experimental systems discussed in the next section.

\section{Experimental Approaches to FT-BQC}
\label{sec:experimental_proposal_for_FTBQC}

\begin{figure}[t]
 \centering
  \includegraphics[width=\columnwidth]{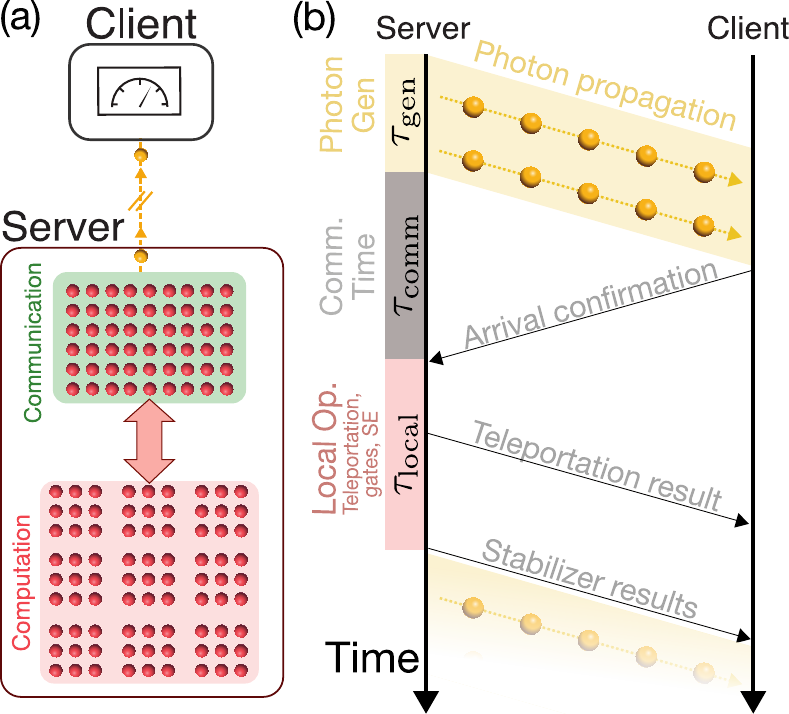}
  \caption{Experimental proposal for the implementation of FT-BQC.
    (a) Schematic of the key elements of an experimental platform for performing FT-BQC: computational qubits (red) hold and manipulate the quantum state, communication qubits (green) interface with the client via a photonic link (yellow), and an interface (red arrow) enables high-fidelity gates between the two elements. 
   (b) Timeline of a FT-BQC. Over time $\tau_{\text{gen}}$, the server generates photons to the client that have been entangled with communication qubits.
   The server waits $\tau_{\text{comm}}$ to receive confirmation of which photons have been measured by the client in order to identify the communications qubits to be used. The server then interfaces the communications with the computation qubits and performs additional local operations (e.g. local gates and syndrome extraction) over a time $\tau_{\text{local}}$.
  }
  \label{fig:exptGen} 
\end{figure}

\ptitle{What are the main features of the experimental implementation of FTBQC: Computation + Communication qubits + fast generation of photonic qubits}
We now analyze the feasibility of our protocol on state-of-the-art quantum hardware.
The envisioned server architecture consists of computational qubits holding and manipulating the quantum state, communication qubits interfacing with the client via photons, and a high-fidelity interface between these two elements [\figref{fig:exptGen}(a)].
This general architecture is compatible with multiple platforms, including neutral atoms~\cite{bluvstein:2024}, trapped ions~\cite{schupp:2021}, superconducting qubits~\cite{osullivan:2024, acharya:2024}, and solid-state spin defects~\cite{wei:2024,knaut:2024, komza:2025, higginbottom:2022, ruskuc:2025, pasini:2024, stas:2022}.

\ptitle{What is the generic framework, regardless of the platform: 
1) Generate photons with communication qubits, 
2) teleport them to the computation qubits, 
3) perform local gates/syndrome extraction}
This proposed architecture enables a simple yet efficient implementation of our fault-tolerant BQC protocol~[\figref{fig:exptGen}(b)].
First, over a time $\tau_{\text{gen}}$, photons entangled with communication qubits are generated and sent to the client.
Due to photon loss, the server must wait for the client's confirmation of a successful photon measurement to identify the usable communication qubits; this requires a communication time $\tau_{\text{comm}}$.
Once a photon measurement is confirmed, the server teleports the corresponding blind gate from the communication to the computational qubit and broadcasts the resulting measurement outcome back to the client [Fig.~\ref{fig:hybrid_rules}].
Finally, the server performs local Clifford gates, logical operations, and syndrome extraction [Fig.~\ref{fig:QEC}(a)], all within a local operation time $\tau_{\text{local}}$.

\begin{figure*}[t]
 \centering
  \includegraphics[width=\textwidth]{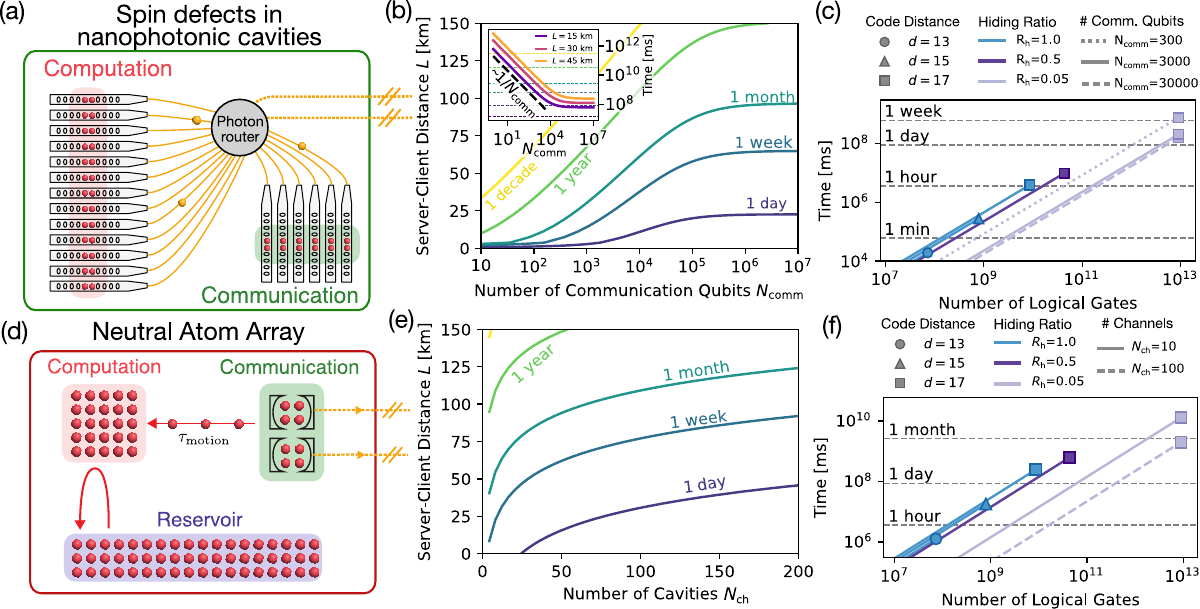}
  \caption{Analysis of the FT-BQC experimental proposal in a neutral atom array platform and in a SiV spin defect network.
  (a) Architecture for a network of SiV spin defects for the execution of FT-BQC. 
  Nanophotonic cavities are divided into a computational group and a communication group.
  The latter is used to generate entanglement with the emitted photons, which is then teleported to the computational qubits upon confirmation of the arrival of the photon.
  The lack of a reservoir of qubits fixes the number of communication qubit making such architecture more susceptible to the server-client distance.
  (b)[(e)] Analysis of the duration of a computation composed of $10^{10}$ logical gates at $R_{\text{h}}=1$, using a surface code of distance $d=17$ using a SiV spin defect platform [neutral atom array].
  (c)[(f)] Analysis of the depth and duration of a blind computation using a SiV spin defect platform [neutral atom array].
  We utilize $L=2~\mathrm{km}$ [$L=10~\mathrm{km}$] and the surface code logical error rates as analyzed in \figref{fig:QEC}.
  We assume the batch execution of $10^3$ blind gates before the client needs to correct the measurement basis based upon the Pauli frame (see Supplementary Materials~\cite{si:} for more details).
  For the SiV platform, we assume there that there are $N_{\text{ch}} = 10$ independent photonic links between the server and the client for additional parallelization.
  (d) Atoms are divided between a computational, a communication, and a reservoir zone. 
  After using a matter qubit to emit a photon, it is moved towards the computation zone while the server awaits a confirmation of its arrival.
  If the measurement is successful, the physical blind gate can be teleported into the computational qubit and the computation can proceed.
  Reservoir qubits are used both for syndrome extraction and to replenish the communication zone.
  }
  \label{fig:exptPlat} 
\end{figure*}

\ptitle{What are the limitations}
Our protocol's performance is bounded by the speed and fidelity of local and non-local operations. 
The protocol's clock speed is limited by the speed of local gates via $\tau_{\text{local}}$.
Additionally, their fidelity dictates the QEC code requirements: 
higher fidelity enables smaller codes,  
reducing the gate count for transversal operations and, subsequently, the number of required photons [Fig.~\ref{fig:QEC}]. 
By contrast, the speed of delegated blind gates is bounded by the photon emission time $\tau_{\text{gen}}$, and 
the server-client distance $L$ through the communication latency $\tau_{\text{comm}} \ge 2L/c$ (where $c$ is the speed of light). 
Practically, photon losses in the link further constrain performance; for a single communication qubit, approximately $1/\eta$ photon attempts are required per physical blind gate. 
Utilizing $N_{\text{comm}}$ communication qubits, however, enables $N_{\text{comm}}$ photonic communication attempts within a single communication cycle $\tau_{\text{comm}}$, significantly increasing performance.

We note that in this architecture, the communication time and need for single photon detection are associated with errors induced by the decoherence of the communication qubit and dark photon detection events.
Fundamentally, such kinds of effects can be incorporated into the communication error $\epsilon_\text{comm}$, and will not dominate in modern quantum platforms~\cite{si:}.

Finally, let us conclude with an important scheduling optimization.
Owing to the way that the Pauli frame affects the measurement angles of the blind operations, large batches of Clifford operations can be implemented within a single communication time, provided the server has access to enough communication qubits.
The key idea is that, for Clifford operations, the client does not need to update their measurement angle based on the Pauli frame, this knowledge is only necessary when implementing non-Clifford operations.
Only at this point must the client receive and process all local measurements, teleportation results, and syndrome extraction to reconstruct the correct Pauli frame.

We now analyze this proposal for two promising experimental platforms for distributed quantum computing: neutral atom arrays and solid-state defects in nanophotonic cavities.

\subsection*{Solid state defects in nanophotonic cavities}

\ptitle{SiV are very good -- strong spin-photon in the optical regime, offering a strong platform for quantum networks and delegated protocols. At the same time, networks of SiV cavities can be brought together to create small quantum computing elements.}
Solid state spin defects~\cite{komza:2025, higginbottom:2022, ruskuc:2025}, and more specifically group IV defects in diamond~\cite{bradac:2019, ruf:2021, knaut:2024, pasini:2024, stas:2022}, are a promising platform for quantum networking and distributed quantum computation applications due to the ability to integrate spin qubits within fabricated nanophotonic cavities, enabling strong spin-photon interactions.

Building upon the promise of the scalable manufacturing of such devices, we envision an architecture composed of a large ensemble of cavities, each hosting an electronic spin defect coupled to a few nearby nuclear spins~[\figref{fig:exptPlat}(a)].
Fast initialization, local operations, and readout can be implemented via the optical link. 
The same optical link enables all-to-all connectivity within the server; connecting these small quantum registers to one another using photonics links via fast and efficient photon routers enables fast and high-fidelity gates across computational qubits and between computation and communication qubits~\cite{simmons:2024a, enthoven:2024, debone:2024}.

Fundamentally, the speed of operations is limited by the electronic spin splitting of the defect, suggesting that local operations and entangled photon emission can be as fast as within a few nanoseconds: $\tau_{\text{local}} \gtrsim 20~\mathrm{ns}$, qubit initialization time $\tau_{\text{init}} \sim 40~\text{ns}$, and photon emission time $\tau_{\text{ph}} \approx 5~\mathrm{ns}$ for Silicon Vacancy (SiV) spin defects, respectively~\cite{wei:2024,knaut:2024}.
Such operating speeds are not yet available, requiring new insights and technological developments, such as addressing ionization limitations on fast photonic gates, as well as the development of new photon generating, routing, and detecting devices~\cite{alexander:2025, pederson:2025}.

Since local operations and photon emission timescales occur at the nanosecond scale, the communication time $\tau_{\text{comm}}$ will be the dominant contribution to the protocol's speed.
For each communication round $\tau_{\text{comm}}$, the server sequentially generates, at most, $N_{\text{comm}}$ entangled photons, before having to wait for the client's measurement confirmation, reducing the platform's rate of execution of logical blind gates.
Note that for large $N_{\text{comm}}$, the generation time can be further reduced by a factor of $1/N_{\text{ch}}$ by multiplexing the communication with the client across $N_{\text{ch}}$ photonic links.

We quantify this expectation by estimating the duration of an algorithm composed of $10^{10}$ blind gates in this platform, as a function of the server-client distance and the number of communication qubits~[\figref{fig:exptPlat}(b)].
Based upon the extensive calculation illustrated in Fig.~\ref{fig:QEC}(h), we estimate requiring a code distance of $d=17$ for $R_{\text{h}}=1$ to achieve such deep circuits.
We observe that increasing the number of communication qubits greatly reduces the duration of the computation, owing to the server's ability to parallelize the photon generation, until the clock speed becomes limited by the speed of photon emission~[\figref{fig:exptPlat}(b,inset)].
To leverage the speed of the platform, it is then critical to either scale up the number of communication qubits, or operate at small distances.

Given these considerations, we now evaluate an SiV-based spin qubit platform for FT-BQC over a distance $L=2~\mathrm{km}$ and utilizing $N_{\text{comm}} = 3000$~[\figref{fig:exptPlat}(c)].
Assuming photon-link-dominated error rates ($\epsilon_{\text{comm}}=0.45\%$, $\epsilon_{\text{local}}=0.045\%$), we estimate computational depth and duration of a blind quantum computation across different surface code distances and hiding fractions of the circuit $R_\text{h}$.
Let us first focus on the role of the code distance.
Based upon our circuit-level QEC simulations, increasing the code distance $d$ enables exponentially longer quantum computations~[Fig.~\ref{fig:QEC}(g)]. 
Leveraging these results, we can estimate the circuit size available for even larger code distances~\cite{si:}.
Crucially, this algorithmic depth improvement induces only a modest increase in the number of photons required per logical blind gate, and thus only a modest increase in the total runtime.

Fixing the code distance, we now consider the effect of modifying the hiding fraction $R_\text{h}$.
Reducing $R_{\text{h}}$ offloads some of the computation to the server, improving the depth and the speed of the computation [\figref{fig:exptPlat}(c, purple)] arising from the improved logical error rates [\figref{fig:QEC}] and the faster application of local non-blind gates, respectively.
This further highlights the importance of carefully choosing $R_{\text{h}}$ to balance the client's requirements of efficiency and security.

Finally, we note that increasing the number of communication qubits remains an important route towards speeding such protocols and reducing the impact of communication time on FT-BQC up until the protocol becomes limited by communication time overhead~[\figref{fig:exptPlat}(c, dotted and dashed)]

\subsection*{Neutral Atoms Arrays}
\ptitle{Introduce the platform: very successful platform for large, high fidelity operations.}
Neutral atom arrays have emerged as powerful quantum computing platforms due to their long coherence times, high-fidelity gates, and intrinsic all-to-all connectivity~\cite{bluvstein:2024, evered:2023, henriet:2020, graham:2022}.
Building upon recent zone-based architectures, we consider a dedicated ``communication zone'' optimized for efficient photon emission~\cite{bluvstein:2024, pattison:2024, sinclair:2024}, \figref{fig:exptPlat}(d).
Recent experiments demonstrated entanglement generation from tweezer-trapped neutral atoms via free-space or cavity-enhanced emission~\cite{welte:2018, vanleent:2022,  young:2022, sinclair:2024,grinkemeyer:2025}. 
Photon emission typically occurs within $\sim100~\mathrm{ns}$, with the photon collection efficiency depending on the emission modality: $\eta_{\text{free}} = 0.114$ and $\eta_{\text{cav}}=0.855$ assuming a photon detection efficiency of $95\%$~\cite{sinclair:2024}.

\ptitle{The usage of different zones brings forth some additional considerations that we will now comment on}
Separating atoms into computation, communication, and reservoir zones offers substantial benefits and distinct challenges. 
A key advantage is the continuous replenishment of pre-initialized atoms, enabling multiple photon emission trials without additional time delays from atom initialization. 
As a result, $\tau_{\text{gen}}$ is primarily limited by the photonic link infrastructure rather than the number of available communication qubits~[\figref{fig:exptPlat}(e)].
More specifically, by segmenting the communication zone into sub-zones corresponding to different photonic links, the photon emission can be parallelized at the expense of an increase in the number of photon-measuring devices at the client's location.
Consequently, the effective photon generation rate is enhanced by the number of photonic links to the clients,  $N_{\text{ch}}$.

On the other hand, atom movement between different zones introduces an additional time scale, $\tau_{\text{motion}}\approx0.5~\mathrm{ms}$ for $0.5~\mathrm{mm}$~\cite{bluvstein:2024}.
Similarly, atom motion sets the natural timescale for local operations and syndrome extraction, $\tau_{\mathrm{local}}\sim 1~\text{ms}$~\cite{bluvstein:2024}.
As such, the speed of blind operations becomes limited by the communication time whenever $L_{\text{motion}} \ge c\tau_{\text{motion}}/2 \approx 75~\mathrm{km}$.

\ptitle{Leveraging these capabilities, we now investigate the capabilities of such a platform for large scale fault tolerant blind quantum computing}
We now evaluate the performance of this platform for a server-client distance of $L=10~\mathrm{km}$~[\figref{fig:exptPlat}(f)].
Assuming again $\epsilon_{\text{comm}}=0.45\%$ and $\epsilon_{\text{local}}=0.045\%$, we observe similar opportunities to the SiV-based platform for performing longer quantum computations by increasing the code distance and decreasing the hiding fraction. 
We conclude by noting that, in analogy to $N_{\text{comm}}$ in the SiV platform, additional photonic links help mitigate the bottleneck of photon generation; for many $N_{\text{ch}}$ the bottleneck then becomes the speed of local operations~[\figref{fig:exptPlat}(b, full vs dashed)].

In summary, neutral atom arrays and SiV spin defect networks each offer unique strengths for implementing FT-BQC. 
Neutral atoms support scalable, highly parallel photon generation suitable for moderate-distance deployment, whereas SiV defects provide rapid local operations favorable at shorter communication distances. 
Future experimental advances and infrastructure optimizations will be critical in realizing the full potential of both platforms for practical FT-BQC implementation.

\section{Discussion and outlook}
\label{sec:conclusion}

\ptitle{conclude main points - big ideas}
In this work, we have introduced a novel approach for scalable fault-tolerant blind quantum computation using a hybrid light-matter architecture. 
By storing the quantum information in matter qubits and using photonic qubits for blind delegated operations, we demonstrate an improvement in the efficiency of delegated computation, which translates into significant gains in the associated error correcting thresholds.
These insights also enable new circuit design opportunities, allowing the client to balance between security and efficiency concerns.

\ptitle{Looking forward: adapting to full photonic + quantum repeaters}
While our work mainly focused on the setting of a matter-qubit based server, the key idea of delegating error correction to the server can also be applied in an all-photonic setting.
In this case, even though the server would generate the entire resource state required for the computation, portions of the state could be immediately measured to implement non-blind local operations (i.e. gates and syndrome extraction), while the client only receives the photons relevant to implementing blind delegated gates.
This would substantially reduce the number of photons transmitted to the client, reducing the requirements for the communication channel and, thus, opening a new avenue towards more robust photonic based computation.
Such ideas could offer new design opportunities for more efficient photonic resource states for delegated computing.

\ptitle{Immediate discussion of our results: Distributed }
At the same time, the ideas explored in this work are also applicable to distributed architectures composed of multiple servers~\cite{wei:2024}.
In such a setting, photon-mediated gates between different servers enable blind (or not-blind) gates to be executed across multiple devices, enabling modular and scalable quantum computing. 
Adapting our ideas to this setting may lead to improved distributed protocols for quantum sensing, communication, and computation.

\ptitle{Field opening questions}

The ability to delegate part of the computation brings forth some important challenges and opportunities in the design of secure delegated algorithms, and their interplay with the underlying error correcting code, and capabilities of the underlying quantum computing platform.

\ptitle{Optimizing Codes for Blind Computation}
In particular, it would be interesting to explore the QEC codes specifically optimized for FT-BQC. 
While our threshold analysis focused primarily on surface codes, the principles we outline naturally extend to all stabilizer codes.
Extending our framework to other classes of QEC codes may enable new opportunities.
Of particular interest is the family of quantum low-density parity-check (qLDPC) codes, which offer asymptotically better scaling in terms of physical-to-logical qubit overhead and could substantially reduce the number of physical qubits and operations needed for blind gates~\cite{tillich:2014, breuckmann:2021}.
Incorporating qLDPC codes into our hybrid architecture may therefore enable deeper circuits with fewer communication qubits required.
Designing such circuits requires jointly considering security, error correction, and efficiency, as these elements are closely interconnected.

\ptitle{More efficient single qubit rotations}
At the same time, developing  effective way of interfacing the photonic qubits with the server's logical qubits is another important avenue of research. 
Although we focused on transversal gates directly acting on the local logical qubits, different strategies can be utilized for implementing blind gates at the logical level, such as e.g. lattice surgery.
Another promising approach would be to perform a blind logical operation on a smaller (or distinct) QEC code, that is then interfaced with the larger computation logical qubit.
Such an approach would reduce the number of required photonic qubits at the expense of a higher error in the delegated gates.
Exploring this trade-off and other strategies might offer even greater efficiency improvements.

\ptitle{What to hide and how?}
Finally, the interplay between security and the design of  blind algorithms should be explored. 
While blindness ensures that no information of the blind operations is leaked to the server, the server's ability to uncover the client's unitary depends on the structure of the computation and the server's understanding of what meaningful computations are of interest to the client.
Recent work has begun to explore this question within a measurement-based setting~\cite{poshtvan:2025}, introducing novel techniques for hiding the structure of the client's calculation---a more general construction that complements the Hamiltonian simulation example presented in Section~\ref{sec:practical_security}.
It remains an open question how to quantify the information leaked when portions of the algorithm are revealed, and how these decisions inform the design of the blind computation. At the same time, this flexibility enables more concrete and efficient computations whereby the notion of security is better understood. While we analyzed the example of Hamiltonian simulation, where selectively choosing which gates to perform blindly allows the client to determine which parts of the computation remain hidden from the server, 
generalizing this remains an open question. It would be interesting to identify the most efficient ways for the client to combine blind and non-blind operations such that the computation is protected to the desired level of security.

\section{Acknowledgments}
We acknowledge insightful discussions with Vladan Vuletić, Aram Harrow, Katherine Van Kirk, Madelyn Cain, Hengyun (Harry) Zhou, Josiah Sinclair, Elham Kashefi, Nishad Maskara, Jake Bringewatt, Yan Qi Huan, Can M. Knaut, Moriz Mertz, Erik E. Knall, Umut Yazlar, Maxim Sirotin, Nazlı Uğur Köylüoğlu, Andi Gu, and Nik O. Gjonbalaj.
G.B.~acknowledges support from the MIT Patrons of Physics Fellows Society.
F.M.~acknowledges support from the NSF through a grant for
ITAMP at Harvard University. 
We acknowledge financial support from Amazon Web Services (award number A50791), 
the National Science Foundation (Grant No. PHY-2012023), 
NSF Engineering Research Center for Quantum Networks (EEC-1941583),
and the Center for Ultra-cold Atoms, an NSF Frontier Center.

\newpage

\appendix

\section{Expressibility}
\label{app:expressibility}

In our manuscript, we describe how the hybrid BQC architecture enables the implementation of both blind and local non-blind gates, allowing greater flexibility in circuit design. 
One advantage of such flexibility is that it provides access to a wider variety of possible unitaries. 
We quantify this argument using expressibility, which can be understood using frame potentials.

We emphasize that the mathematical definition of expressibility differs between works; the formalism introduced in this section is ours.

\subsection*{Our Characterization of Expressibility}

In our work, we compare the expressibility of a partially-blind circuit (such as the $\theta$-blind Pauli rotations) to the expressibility of a fully-blind circuit (such as the fully-blind bricklayer), and ask how many blind gates each circuit requires to achieve the same degree of expressiblity. We assume the rotation angles $\{\theta\}$ of the blind gates may be modeled as random variables following a uniform random distribution. 
Fully- or partially-blind circuits may then be thought of as parameterized circuits, and the ensemble of unitaries associated with a circuit is the distribution of unitaries parameterized by its uniformly random blind rotation angles. If the $k$th frame potentials of two circuits are approximately equal, then we say the two circuits have approximately equal expressiblity for degree-$k$.

Given that frame potentials may be modeled with exponential decay [Fig.~\ref{fig:express_app}(a,b)], we can derive a general form relating circuits with equal expressibility. Frame potentials converge to the form ~\cite{liu:2022},
\begin{align}
    \sqrt{\mathcal{F}_{\mathcal{E}_i}^{(k)}- \mathcal{F}_\text{Haar}^{(k)}} = A_{\mathcal{E}_i}e^{-l_i/C_{\mathcal{E}_i}}
    \label{eqn:frame_potential_exponential_decay}
\end{align}
where $l_i$ is the number of layers, or the depth, of in circuit $i$. The other variables $A_{\mathcal{E}_i}$ and $C_{\mathcal{E}_i}$ are constants with respect to $l_i$ and for a given $k$, $n$, and $\mathcal{E}_i$.

Setting the frame potential of an ensemble $\mathcal{E}_1$ equal to that of another ensemble $\mathcal{E}_2$,
\begin{align}
    l_1 = \frac{C_{\mathcal{E}_1}}{C_{\mathcal{E}_2}} l_2 - C_{\mathcal{E}_1} \ln\bigg(\frac{A_{\mathcal{E}_2}}{A_{\mathcal{E}_1}}\bigg),
\end{align}
and the number of layers required for given ansatz to achieve equal expressibility grows in proportion to that of other ansatze.

If the number of blind gates per circuit layer is constant in $n$ for $\mathcal{E}_1$ (as it is in the $\theta$-blind Pauli rotations) and linear in $n$ for $\mathcal{E}_2$ (as it is in the fully-blind bricklayer), then
\begin{align}
    N_1 = r N_2 - C_{\mathcal{E}_1} \ln\bigg(\frac{A_{\mathcal{E}_2}}{A_{\mathcal{E}_1}}\bigg),
\end{align}
where $r \sim \frac{C_{\mathcal{E}_1}}{C_{\mathcal{E}_2}} \cdot \frac{1}{n}$. We note that $A_{\mathcal{E}_i}$ and $C_{\mathcal{E}_i}$ do depend on $n$ and $k$, but $A_{\mathcal{E}_i}$ is not exponential and $C_{\mathcal{E}_i}$ is sublinear~\cite{liu:2022}. This scaling is reflected in Fig.~\ref{fig:Fig_circ_design_express}.

\subsection*{Numerical Methods}

\begin{figure}[t]
 \centering
  \includegraphics[width=\columnwidth]{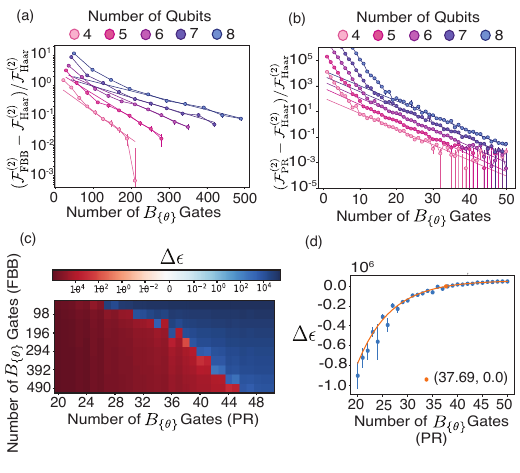}
  \caption{Intermediate plots for expressibility. (a) Frame potential for fully-blind bricklayer (FBB) circuits of varying depths, and thus a varying number of blind gates. (b) Frame potential for $\theta$-blind Pauli rotation (PR) circuits of varying depths, and thus a varying number of blind gates. (c) A 2D histogram of difference in frame potential (captured by $\Delta \epsilon$) between FBB and PR circuits for $n=8$ qubits. (d) Difference in frame potential (captured by $\Delta \epsilon$) between a FBB circuit with 8 qubits and 196 blind gates, and PR circuits of varying depths.
  }
  \label{fig:express_app} 
\end{figure}

To evaluate the $k$th frame potential for a given circuit, we sample uniformly at random two unitaries that may be implemented by the circuit and evaluate $|\text{Tr}(U^\dagger V)|^{2k}$. We repeat this enough times to generate a distribution. The average of this distribution is an estimate on the frame potential. We chose $k=2$ because it is the smallest value of $k$ yielding non-trivial results, as 1-designs are easy to achieve and most of our circuits already form 1-designs with a frame potential of 1.

We first evaluate the frame potential of fully-blind bricklayer (FBB) circuits, denoted $\mathcal{F}^{(2)}_\text{FBB}$, for various depths. This is done with $n=4,5,6,7,8$ qubits, and the frame potential is plotted with respect to the number of $B_{\{\theta\}}$ gates required for each circuit depth [Fig.~\ref{fig:express_app}(a)]. This is repeated for $\theta$-blind Pauli rotation (PR) circuits, where each layer in the circuit is a single $n$-qubit $\theta$-blind Pauli rotation [Fig.~\ref{fig:express_app}(b)]. The frame potential for PR circuits is denoted $\mathcal{F}^{(2)}_\text{PR}$.

For each value of $n$, a 2D histogram of 
\begin{align}
    \Delta \epsilon \equiv d^{k} \sqrt{F_\text{FBB}^{(k)}-F_\text{Haar}^{(k)}}-d^{k}\sqrt{F_\text{Haar}^{(k)}-F_\text{Haar}^{(k)}}
\end{align}
for varying depths of FBB and PR circuits is generated [Fig.~\ref{fig:express_app}(c)]. We want to find the points on the histogram where the frame potentials of both circuits are equal and $\Delta \epsilon = 0$. For each row, i.e., for a fixed depth of the FBB circuit, we plot $\Delta \epsilon$ for varying PR circuit depths and identify the number of $B_{\{\theta\}}$ gates in a PR circuit that achieves $\mathcal{F}^{(2)}_\text{FBB} = \mathcal{F}^{(2)}_\text{PR}$ [Fig.~\ref{fig:express_app}(d)]. These values form the data shown in Fig.~\ref{fig:Fig_circ_design_express}.

\section{QEC Simulations}
\label{app:QEC_simulations}

\subsection*{Methods}
We conduct circuit-level numerical simulations to assess the performance of blind logical algorithms under realistic noise conditions. Our simulations focus on the rotated surface code at distances \( d = 3,5,7,9 \). 
Logical error results with larger distances used in Section~\ref{sec:experimental_proposal_for_FTBQC} are achieved by fitting.
The logical circuits consist of 10 or 20 layers, each acting on four logical qubits. Each layer applies a combination of randomly chosen single-qubit logical gates, randomly selected pairs of two-qubit transversal logical gates, and periodic syndrome extraction rounds. A detailed description of the noise model applied to each gate type is provided in the Supplementary Materials~\cite{si:}.

Simulations are performed using the open-source Clifford circuit simulator Stim~\cite{gidney:2021}. Since our circuits consist entirely of Clifford operations and a Pauli noise model, they can be efficiently simulated in polynomial time with respect to the number of physical qubits, as guaranteed by the Gottesman-Knill theorem~\cite{gottesman:1997}. 

To estimate the logical error rate, we employ Monte Carlo sampling of physical error configurations based on the noise model. For each sample, the corresponding syndrome measurements and logical observable measurements are generated. The resulting syndromes are decoded using the Most-Likely Error (MLE) correlated decoder, which accounts for correlated errors between logical qubits, as introduced in~\cite{cain:2024} and further applied in~\cite{baranes:2025, zhou:2024}. A logical error is recorded if the final decoded logical measurement deviates from the expected noiseless outcome.

\subsection*{Logical Error Characterization}

\begin{figure}[t]
    \centering    
    \includegraphics[width=\columnwidth]{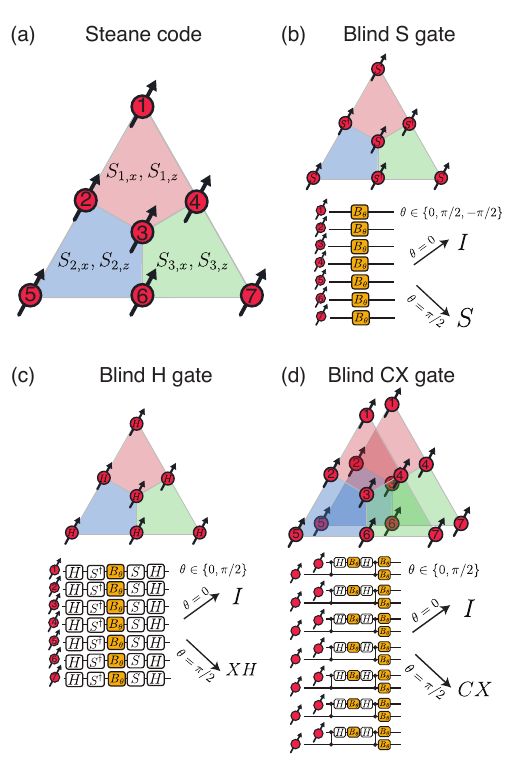}
    \caption{Example of blind gates in the Steane code. (a) Steane code data qubits and stabilizer definitions. $S_{i,x}$ and $S_{i,z}$ (for $i = 1,2,3$) are defined as products of $X$ or $Z$ operators, respectively, on neighboring data qubits. (b) Blind $S$ gate. (c) Blind $H$ gate. (d) Blind $CX$ gate.
    }
    \label{fig:Steane_example}
\end{figure}

To quantify the impact of noise on logical circuits, we define the algorithmic logical error rate, which ranges from \( 0 \) to \( p_{L,\text{max}} = 1 - 1/{2^{N_{\text{q}}}} \), where \( p_{L,\text{max}} \) corresponds to the logical error rate of a maximally mixed state with \( N_{\text{q}} \) logical qubits.

The logical error per round is given by:
\begin{equation}
    p_{L,\text{round}} \approx \frac{p_L}{N_{\text{layers}} \cdot N_{\text{rounds}}},
\end{equation}
where \( N_{\text{layers}} \) is the number of gate layers and \( N_{\text{rounds}} \) is the number of syndrome extraction rounds per layer.

The logical error per gate follows:
\begin{equation}
    p_{L,\text{gate}} = 1 - (1 - p_{L,\text{round}})^{\frac{1}{N_{\text{gpl}}}},
\end{equation}
where \( N_{\text{gpl}} \) represents the expected number of gates per layer.
The final logical fidelity after executing \( N_{\text{tot}} \) is:
\begin{equation}
    F_{\text{L}} \approx (1 - p_{L,\text{gate}})^{N_{\text{total gates}}},
\end{equation}
with the total number of logical gates scaling as:
\begin{equation}
    N_{\text{tot}} = N_{\text{gpl}} \cdot N_{\text{layers}} \cdot N_{\text{rounds}}.
\end{equation}

Rearranging, the maximum number of logical gates achievable while maintaining a final fidelity \( F_L \) is:
\begin{equation}
    N_{\text{tot}} \approx \frac{\log{F_{\text{L}}}}{\log{(1 - p_{L,\text{gate}})}}.
\end{equation}

These equations provide a direct method for estimating computational depth in blind logical circuits. A detailed numerical validation is presented in the Supplementary Materials~\cite{si:}.

\section{Blind logical gates example on the Steane code}
\label{app:Steane_code}
Here we show a concrete example of logical blind gates on the Steane code $[[7,1,3]]$ (see Appendix Fig.~\ref{fig:Steane_example}). 
A transversal blind $S$ gate can be realized by applying $B_{\{\theta\}}$ rotations with angles $\theta \in \{0,\pi/2,-\pi/2\}$ on each data qubit, producing either the identity or $S$. 
A blind Hadamard gate ($H$) can similarly be implemented using local Clifford operations (e.g., $U=HS$) around blind rotations with angles $\theta \in \{0, \pi/2\}$, producing either the identity or $H$. 
A transversal $CX$ gate involves global blind $CX$ blocks, combining blind rotations and local $CZ$ gates across corresponding data qubits. 
Crucially, according to Section~\ref{sec:hybrid_rules_of_the_game}, all of these implementations require only one successful photon measurement per qubit, ensuring inherent loss tolerance.

\bibliography{references2}

\end{document}

\definecolor{mygray}{gray}{0.6}
\newcommand\numberthis{\addtocounter{equation}{1}\tag{\theequation}}
\newcommand{\insertimage}[1]{\includegraphics[valign=c,width=0.04\columnwidth]{#1}}

\newcommand{\fm}[1]{\textcolor{blue}{\small FM:#1}}
\newcommand{\GB}[1]{{[\small \color{blue}GB: #1]}}
\newcommand{\IW}[1]{{[\small \color{red}IW: #1]}}
\newcommand{\rem}[1]{{\color{red}\sout{#1}}}
\newcommand{\note}[1]{\textcolor{blue}{#1}}

\date{\today}
\setcounter{page}{1}
\renewcommand{\thepage}{S\arabic{page}}
\renewcommand{\thesection}{S\arabic{section}}
\renewcommand{\theequation}{S\arabic{equation}}
\renewcommand{\thefigure}{S\arabic{figure}}
\renewcommand{\thetable}{S\arabic{table}}
\setcounter{section}{0}
\def\lc{\left\lceil}   
\def\rc{\right\rceil}
\def\lf{\left\lfloor}
\def\rf{\right\rfloor}

\title{Supplementary Materials of ``Designing Fault-Tolerant Blind Quantum Computation"}
\date{\today}
\maketitle
\tableofcontents
\newpage

\maketitle

\section{Blindness Definition}
The client selects the measurement angles $\phi$ and is the only party aware of the measurement results $s = \pm 1$. Given a $+1$ result, a Pauli correction $P$ must be applied, which is known only to the client. This ensures blindness, as the server observes a density matrix that sums over all possible measurement outcomes, preventing it from inferring the measurement angles or results.

To prove that a gate is blind, it is necessary to show that the server's local density matrix, $\rho_{\text{final}}$, after the gate implementation $U({\phi}, {s})$, does not depend on the measurement angles ${\phi}_{i=1}^n$. 
The server's final density matrix can be written as:
\begin{equation} \rho_{\text{final}} = \frac{1}{2^n} \sum_{j=1}^{2^n} P_j U({\phi}) \rho_j U^\dagger({\phi}) P_j^\dagger, \end{equation}
where $P_j$ represents the Pauli correction corresponding to the measurement result $s_j$ over $n$ measurement outcomes.

\section{Implementing $B_\theta$ Gates}
\label{SI:implementing_B_theta_gates}

Assuming the use of quantum memories, the procedure to implement a $B_\theta$ gate is the following.

\begin{enumerate}
    \item The server generates a bell pair between a photon ($\gamma$) and a communication qubit, yielding the state $(\ket{0}_\gamma \otimes \ket{0}_\text{comm} + \ket{1}_\gamma \otimes \ket{1}_\text{comm})/\sqrt{2}$. The specific process used to create this bell pair depends on the server's quantum computing platform.
    \item The resulting entangled photon is sent to the client, who attempts to measure the photon in the basis $\ket{+}_\theta = (\ket{0} + e^{i\theta} \ket{1})_\gamma/\sqrt{2}$ and $\ket{-}_\theta = (\ket{0} - e^{i\theta} \ket{1})_\gamma/\sqrt{2}$ with measurement outcome $s = 0$ and $1$ respectively. Steps 1 and 2 are repeated with new photons are communication qubits until a photon detection by the client heralds a successful measurement. The communication qubit is now in the state $(\ket{0} + (-1)^se^{i\theta}\ket{1})_\text{comm}/\sqrt{2}$, effectively representing a $Z^s R_z(\theta)$ gate.
    \item A local CNOT on the communication qubit and conditioned on a computation qubit in the state $(a \ket{0} + b \ket{1})_\text{mq}$ is performed, yielding $$\left[\ket{0}_\text{comm}(a\ket{0} + b(-1)^se^{i\theta} \ket{1})_\text{comp} + \ket{1}_\text{comm}(a(-1)^se^{i\theta} + b \ket{1})_\text{comp}\right]/\sqrt{2}.$$ The communication qubit is measured in the $z$ basis with measurement outcome $m=0,1$, such that the computation qubit is in the state $(a \ket{0} + b(-1)^s e^{(-1)^m i \theta} \ket{1})_\text{comp}$. The gate $Z^sR_z((-1)^m\theta)$ is transferred to the computation qubit.
\end{enumerate}

If the measurement outcome of the communication qubit is $m=0$, then we have successfully implemented a $B_\theta$ gate. However, if $m=1$ a rotation by $-\theta$ is implemented, and we correct this by repeating steps 1-3 with rotation angle $2\theta$ instead of $\theta$. Should $m = 0$ then $-\theta + 2\theta = \theta$ and we achieve the desired rotation, otherwise if $m = 1$ we must repeat this process recursively.

By discretizing possible values of $\theta$ such that $\theta \in \{2\pi j / 2^c\}$ where $j = 0, 1, \ldots, 2^c-1$ and $c$ is an integer, we can show that steps 1-3 must be repeated at most $c$ times. If with any communication qubit measurement $m = 0$, the desired rotation is achieved. If communication qubit measurement yields $m=1$ for each repetition, then on the $c$th measurement the rotation angle becomes
\begin{align}
    -\theta-2\theta-4\theta-\ldots-2^{c-1}\theta = -\theta\sum_{x=0}^{c-1}2^x = (1-2^c)\theta = \theta - 2\pi j
\end{align}
and the desired rotation is achieved. On average, these steps are expected to be repeated
\begin{align}
    \sum_{j=1}^{c-1} \frac{j}{2^j} +\frac{c}{2^{c-1}} = 2-2^{1-c}
\end{align}
times.

Throughout these repetitions blindness is preserved, as the server must notify the client of the outcome of $m$ but the client does not provide additional information to the server. The implemented gate is $Z^{s}R_z(\theta)$, where $s = \left(\sum_j s_j\right) \mod 2$. The sum is over all performed repetitions of steps 1-3 and $s_j$ is the photon measurement outcome on the $j$th repetition. Since all $s_j$ are held by the client, the Pauli correction $Z^{\sum_j s_j}$ still results in blindness.

\hspace{5mm}

Without quantum memories, steps 1 and 2 can be performed by directly entangling photons with qubits storing the quantum state of the computation. However, this requires all photons in a computation to be measured successfully, and a computation must be repeatedly attempted until all measurements are successful.

\section{Brickwork and universal blind unit cell}
The standard implementation of blind quantum computation employs a resource state known as the brickwork state~\cite{fitzsimons:2017, broadbent:2009}. The brickwork state is composed of multiple unit cells, each capable of implementing a universal blind gate over two neighboring qubits. The traditional implementation requires eight photons.

\subsection{Hybrid universal unit cell}
\begin{figure}
    \centering
    \includegraphics[width=1\linewidth, trim=0cm 0cm 0cm 0cm]{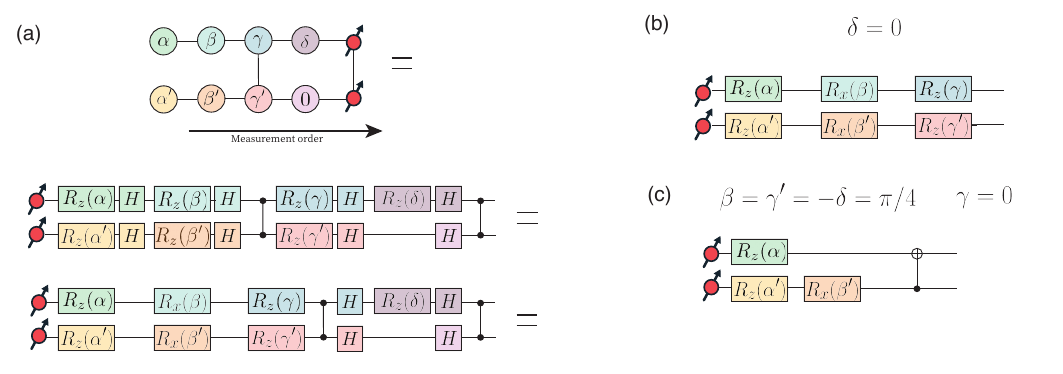}
    \caption{A general brickwork state utilizing only seven photons for blind universal quantum computation. (a) The brickwork structure is presented, with one node consistently measured in the $0$ phase (X basis). This cluster state is equivalent to the circuits shown below. (b) On the left, setting $\delta = 0$ results in two universal single-qubit gates. On the right, choosing $\beta = \gamma' = -\delta = \pi/4$ and $\gamma = 0$ yields a CNOT gate with additional blind rotations. This brickwork configuration, with the constraint of a node always measured in the $0$ phase, can be used to conceal the execution of single-qubit or two-qubit gates. It also allows for universal single-qubit gates. In our protocol, this is achieved using only seven photons instead of eight.}
    \label{fig:brickwork_7_photons}
\end{figure}

By examining the circuit in Fig.~\ref{fig:brickwork_7_photons}, it becomes apparent that only seven phases are required to generate universal gates over two qubits, in contrast to the eight phases needed for the fully photonic implementation. This observation suggests that, in a hybrid protocol where the Hadamard (H) and controlled-Z (CZ) gates are applied by the server, only seven photons are necessary to construct the standard brickwork state.

The unitary operation applied, as a function of the measurement angles ${i=\alpha,\alpha',\beta,\beta',\gamma,\gamma',\delta}$ and the measurement results $\{s_i\}$, is given by:
\begin{equation}
    CZ H_1 H_2 Z_1^{s_\delta} R_{z1}(\delta) H_1 H_2 Z_1^{s_\gamma} R_{z1}(\gamma) Z_2^{s_{\gamma'}} R_{z2}(\gamma') CZ H_1 H_2 Z_1^{s_\beta} Z_2^{s_\beta'} R_{z1}(\beta) R_{z2}(\beta') H_1 H_2 Z_1^{s_\alpha} Z_2^{s_\alpha'} R_{z1}^{\alpha} R_{z1}^{\alpha'}
\end{equation}

For $\delta = 0$, this unitary implements two blind, single-qubit universal gates, controlled by the phases $\alpha, \beta, \gamma$ for qubit $q_1$ and $\alpha', \beta', \gamma'$ for qubit $q_2$.

For $\beta = \gamma' = -\delta = \pi/4$ and $\gamma = 0$, the unitary implements a two-qubit controlled-NOT (CX) gate with additional blind rotations controlled by $\alpha, \alpha', \beta'$.

\section{Decomposing the universal cell into smaller cells}
The hybrid implementation introduces greater flexibility, allowing the decomposition of the standard universal cell into smaller building blocks.

\subsection{Blind universal single qubit gate}

The unitary applied, as a function of the measurement angles $\{ \phi \}_{i=1}^{3}$
and the measurement results $\{s \}_{i=1}^{3}$ is:
\begin{equation}
    U_B(\{\phi\},\{s\}) = Z^{s_3}R_z(\phi_3)HZ^{s_2}R_z(\phi_2)HZ^{s_1}R_z(\phi_1)
\end{equation}

Using the identities $HX^s = Z^s H$, $HR_z(\phi)H=R_x(\phi)$, and $R_z(\phi)X^s = X^s R_z((-1)^s \phi)$, we obtain:
\begin{equation}
    U_B(\{\phi\},\{s\}) = Z^{s_3} X^{s_2} Z^{s_1}
    R_z((-1)^{s_2} \phi_3)R_x((-1)^{s_1} \phi_2) R_z(\phi_1)
\end{equation}

\subsection{Blind two qubit gate}
Consider two qubits, $q_1$ and $q_2$. The unitary applied, as a function of the measurement angle $\phi$ and the measurement result $s$, is:
\begin{equation}
    U_C(\phi, s) = CZ H_1 X_1^s R_{z1}(\phi) H_1 CZ = X_1^s X_2^s CZ H_1 R_z(\phi) H_1 CZ
\end{equation}

To prove the blindness of this block, note that depending on the photonic measurement result, there are two possible unitaries:
\begin{equation} 
U_1 = CZ , R_{x1}(\phi) CZ, 
\end{equation}
\begin{equation} 
U_2 = CZ , R_{x1}(\phi + \pi) CZ. 
\end{equation}

Given an initial state $\rho$, the density matrix on the server's side after the gate application is:
\begin{equation}
        \rho_{\text{final}} = \frac{1}{2} (U_1^\dagger \rho U_1 + U_2^\dagger \rho U_2) = \\
          \frac{1}{2} ((CZ R_{x1}(\phi) CZ) \rho (CZ R_{x1}(\phi) CZ) + (CZ R_{x1}(\phi+\pi) CZ) \rho (CZ R_{x1}(\phi+\pi) CZ)),
\end{equation}

which, after applying gate identities, is independent of the phase $\phi$, confirming the blindness of the protocol.

\newpage

\section{Fidelity and Efficiency Simulations}
This section describes the simulations conducted to estimate the fidelity, efficiency, and unitary space of blind quantum algorithms in our proposed framework.

We consider a random brickwork state, where each layer is constructed using single-qubit blind blocks and entangling blocks.
We vary the ratio of blind blocks, denoted as $R_\text{h}$. For a given ratio $R_\text{h}$, the fraction $R_\text{h}$ of the bricks are blind, while the remaining $1-R_\text{h}$ bricks are not blind. The non-blind bricks are represented by random local gates, with phases chosen uniformly at random. We assume these gates have perfect fidelity and efficiency since they are executed locally on the server's quantum computer.
The single-qubit brick uses three $B_{\{\theta\}}$ gates, while the entangling brick uses a single $B_{\{\theta\}}$ gate.
The simulations are carried out using Monte Carlo sampling. For each shot, we randomly choose the specific blind bricks and phases.

\subsection{Efficiency Calculation}

To implement a blind rotation, the server entangles a photon with a matter qubit and sends the photon to the client, after which the photon may or may not be received and measured successfully. If the photon is lost and not measured successfully, the server must try again with another photon. For a given blind circuit, let $N_\gamma$ be the expected number of photons required to successfully implement it.

Assuming the use of quantum memories, the server may continuously send photons until he receives a ``click'' from the client indicating a successful measurement. Upon this ``click'' the blind gate is transferred to the desired matter qubit, implementing the desired gate. This ``click'' is known as heralding. The use of quantum memories enables the loss-tolerant implementation of $B_{\{\theta\}}$ described in~\ref{SI:implementing_B_theta_gates}

With quantum memories, $N_\gamma$ is the expected number of photons required to implement a single loss-tolerant $B_{\{\theta\}}$ gate, $(2-2^{1-c})/\eta$, times the number of $B_{\{\theta\}}$ gates $N$:
\begin{align}
    N_\gamma = \frac{N(2-2^{1-c})}{\eta}.
\end{align}

Without quantum memories but with heralding, a circuit is run only when a sequence of $N$ photons are all measured successfully. When a photon in this sequence is not measured successfully, the circuit must be stopped and restarted. Every time the circuit is restarted, we call the resulting sequence of photons a new attempt. An attempt succeeds with probability $\eta^N$, using $N$ photons. An attempt fails with probability $(1-\eta^N)$, using on average
\begin{align}
    N_\text{fail} = \frac{1}{1-\eta^N}\sum_{n=1}^N \eta^{n-1} (1-\eta) n = \frac{1}{1-\eta} + \frac{N\eta^N}{1-\eta^N}
\end{align}
photons. In addition, the circuit must be restarted and another attempt made. The total number of expected photons required to implement the circuit of $N$ blind $B_{\{\theta\}}$ gates is therefore defined by
\begin{align}
    N_\gamma = \eta^N N + (1-\eta^N) (N_\gamma + N_\text{fail}).
\end{align}
Solving for $N_\gamma$ and in the limit of small $\eta$,
\begin{align*}
    N_\gamma \approx \frac{N_\text{fail}} {\eta^N} \approx \frac{1} {\eta^N(1-\eta)}
\end{align*}

We define the total \emph{circuit efficiency} as $\eta_\text{circ} = 1/N_\gamma$.

\subsection{Fidelity Calculation}

We assume that each $B_{\{\theta\}}$ gate introduces a depolarizing channel on the matter qubit involved in the gate. This is represented by the following transformation:

\begin{equation}
    \rho \rightarrow (1 - \epsilon_{B_{\{\theta}\}}) \rho + \frac{\epsilon_{B_{\{\theta\}}}}{3} (X\rho X + Y\rho Y + Z\rho Z)
\end{equation}

where $\epsilon_{B_{\theta}}$ is the $B_{\theta}$ gate infidelity, connected to the single photon gate infidelity by the expected number of photons required to implement a loss-tolerant gate (Section~\ref{SI:implementing_B_theta_gates}), $\epsilon_{B_{\theta}} = (2-2^{1-c} )\cdot \epsilon $ with $\epsilon$ the single photon error rate.

Additionally, we assume that each 2-qubit gate introduces a 2-qubit depolarizing channel, represented by

\begin{align}
    \rho \rightarrow (1 - \epsilon_\text{2Q}) \rho + \frac{\epsilon_\text{2Q}}{15} \sum_P P\rho P
\end{align}

where $\epsilon_\text{2Q}$ is the 2-qubit gate infidelity and the sum is over all Pauli strings of lengths 1 and 2.

For a given Monte Carlo shot, we propagate all errors through the quantum circuit to obtain the final density matrix $\rho_{\text{noise}}$. We then generate another circuit without any noise to obtain the ideal density matrix $\rho_{\text{ideal}}$. The algorithmic fidelity is computed as:

\begin{equation}
    F_{\text{algorithm}} = \left( \text{tr} \left( \sqrt{\sqrt{\rho_{\text{noise}}} \rho_{\text{ideal}} \sqrt{\rho_{\text{noise}}}} \right) \right)^2
\end{equation}

For a circuit of $n$ qubits, the minimum possible fidelity is $1/2^n$, corresponding to complete decoherence into the maximally mixed state. 

\vspace{5mm}

We now provide a lower bound on the fidelity of the partially blind circuit by concatenating the effect of many noisy gates.
The key observation is that each depolarization channel with error $\epsilon$ reduces the fidelity of by, at most, a factor of $1-\epsilon = f$. 
The final fidelity can be quickly lower bounded by counting the number of single and two qubit depolarizing channels associated with delegated and local gates, respectively:
\begin{equation}
    F \ge f_{\text{1Q}}^{N_{1Q}}f_{\text{2Q}}^{N_{2Q}}
\end{equation}

We now focus on the particular circuit construction considered in our work. 
For the blind operations, each single qubit gate necessitates three blind qubit rotations and the error is modeled by three applications of the single qubit depolarization channel.
The two qubit blind gate, requires 2 two qubit operations, and a single qubit blind operation.
The corresponding loss of fidelity is then $f_{comm}^3$ and $f_{CZ}^2 f_{comm}$, respectively.

For the local gates, we assume that the single qubit operations are perfect, and that the two qubit operations have error.
Because we are revealing the gate, the structure of the two qubit blind gate can be replaced by a single two-qubit gate, inducing a loss of fidelity $f_{CZ}$.

For some value of $R_h$, the number of blind single qubit gates will be, $\frac{2N}{3}R_h$ where $N$ is the total number of gates, the number of blind two qubit gates will be $\frac{N}{3}R_h$, and the number of local two-qubit gates will be $\frac{N}{3}(1-R_h)$.
The fidelity $F$ of the entire circuit can then be estimated as:
\begin{align}
F &= \left[f_{comm}^3\right]^{\frac{2N}{3}R_h} \left[ f_{CZ}^2 f_{comm} \right]^{\frac{N}{3}R_h} \left[ f_{CZ} \right]^{\frac{N}{3}(1-R_h)} \\
&= \left[  \frac{f_{comm}^{2+1/3} f_{CZ}^{2/3}}{f_{CZ}^{1/3}} \right]^{NR_h} \left[f_{CZ}^{1/3}\right]^{N} \\
&= \left[\frac{f_{\text{blind}}}{f_{\text{local}}} \right]^{N R_h} f_{\text{local}}^N \label{eq:Scaling}
\end{align}
where $f_\text{blind} = f_{comm}^{2+1/3} f_{CZ}^{2/3}$ and $f_\text{local} = f_{CZ}$ encode the fidelity of local and blind gates, respectively.

\subsection{Number of Unitaries Calculation}

We estimate the number of distinct unitaries possible for a given blind circuit to understand the size of the unitary space that the server must consider to guess the client’s intended unitary. This is given by: $|\{\theta\}|^N = 2^{cN}$. Then, the probability that a server randomly guesses the client's computations is $1/|\{\theta\}|^N $. A similar counting argument is made in~\cite{poshtvan:2025} to derive a lower bound on the required communication overhead.

\section{Expressibility Background and Additional Numerics}

The notion of expressibility has been previously used in studies of Variational Quantum Algorithms (VQAs), which is a hybrid quantum-classical framework. 
It is typically used to quantify how well a circuit ansatz can explore the space of possible solutions~\cite{sim:2019,nakaji:2021}. 
In the context of BQC, we use expressibility to quantify the breadth of the unitary space accessible to the client, given a circuit with a specified number of blind gates.

Expressibility is often studied through Haar random $k$-designs~\cite{sim:2019, nakaji:2021, hunter-jones:2019, liu:2022}. 
The Haar measure corresponds to the uniform distribution across the space of all possible unitaries or quantum states, maximizing expressibility. 
This continuous distribution requires an exponential-depth circuit to exactly implement, and is not easily compared with discrete distributions. 
To study the randomness of a distribution of unitaries, $k$-designs are thus often used. 

A discrete ensemble of unitaries $\mathcal{E}$ with cardinality $|\mathcal{E}|$ is a \textit{unitary $k$-design} if the $k$th moment of $\mathcal{E}$ is equal to the $k$th moment of the Haar measure, i.e., $\hat\Phi_\mathcal{E}^{(k)} = \hat\Phi_\text{Haar}^{(k)}$.
Here, $\hat\Phi_\mathcal{E}^{(k)}$ and $\hat\Phi_\text{Haar}^{(k)}$ are moment operators, defined as
\begin{align}
    \hat\Phi_\mathcal{E}^{(k)} = \frac{1}{|\mathcal{E}|}\sum_{U \in \mathcal{E}} U^{\otimes k} \otimes (U^\dagger)^{\otimes k}
\end{align}
for the discrete set $\mathcal{E}$ and
\begin{align}
    \hat\Phi_\text{Haar}^{(k)} = \int_{U \in \text{Haar}} dU \, U^{\otimes k} \otimes (U^\dagger)^{\otimes k}
\end{align}
for the continuous Haar distribution.

An alternative and equivalent definition of a $k$-design can be made defining twirling operators $\Phi_\mathcal{E}^{(k)}(\mathcal{O})$ and $\Phi_\text{Haar}^{(k)}(\mathcal{O})$, where
\begin{align}
    \Phi_\mathcal{E}^{(k)}(\mathcal{O}) = \frac{1}{|\mathcal{E}|}\sum_{U \in \mathcal{E}} U^{\otimes k} (\mathcal{O}) (U^\dagger)^{\otimes k}
\end{align}
for the discrete set $\mathcal{E}$ and
\begin{align}
    \Phi_\text{Haar}^{(k)}(\mathcal{O}) = \int_{U \in \text{Haar}} dU \, U^{\otimes k} (\mathcal{O}) (U^\dagger)^{\otimes k}
\end{align}
for the continuous Haar distribution.
Then, a $k$-design is an ensemble satisfying $\Phi_\mathcal{E}^{(k)}(\mathcal{O})= \Phi_\text{Haar}^{(k)}(\mathcal{O})$.
Designs with high $k$ are pseudorandom, while low $k$ indicates a highly structured ensemble. 

To quantify the closeness of a unitary ensemble to a $k$-design, $\mathcal{E}$ is called an \textit{$\epsilon$-approximate $k$-design} if the diamond distance between the two channels is less than or equal to $\epsilon$, i.e.,
\begin{align}
    ||\Phi_\mathcal{E}^{(k)}(\mathcal{O})-\Phi_\text{Haar}^{(k)}(\mathcal{O})||_\diamond \leq \epsilon.
    \label{eqn:epsilon_approx_k_design}
\end{align}

The diamond norm can be reduced to a semidefinite program and is therefore not numerically-friendly to calculate; the \textit{frame potential} simpler to estimate. Frame potential is defined as
\begin{align}
    \mathcal{F}_\mathcal{E}^{(k)} = \int_{U,V \in \mathcal{E}} dU \, dV\, |\text{Tr}(U^\dagger V)|^{2k}
\end{align}
for an ensemble of unitaries $\mathcal{E}$, and satisfies
\begin{align}
    ||\hat\Phi_\mathcal{E}^{(k)}-\hat\Phi_\text{Haar}^{(k)}||_\text{F}^2 = (\mathcal{F}_\mathcal{E}^{(k)} -\mathcal{F}_\text{Haar}^{(k)})
\end{align}
where $||\cdot||_\text{F}$ is the Frobenius norm~\cite{hunter-jones:2019}. This is used to bound Eq. \eqref{eqn:epsilon_approx_k_design},
\begin{align}
    ||\Phi_\mathcal{E}^{(k)}(\mathcal{O})-\Phi_\text{Haar}^{(k)}(\mathcal{O})||_\diamond^2 & \leq d^{2k}||\hat\Phi_\mathcal{E}^{(k)}-\hat\Phi_\text{Haar}^{(k)}||_\text{F}^2 \\
    & = d^{2k} (\mathcal{F}_\mathcal{E}^{(k)} -\mathcal{F}_\text{Haar}^{(k)})
\end{align}
where $d$ is the Hilbert space dimension, $\mathcal{F}_\text{Haar}^{(k)} = k!$, and $\mathcal{F}_\mathcal{E}^{(k)} \geq \mathcal{F}_\text{Haar}^{(k)}$. The smaller the frame potential of an ensemble, the closer that ensemble is to uniform Haar randomness. Thus, two ensembles with equal frame potential may be considered as equidistant from a $k$-design. 

\begin{figure}[t]
 \centering
  \includegraphics[width=0.5\columnwidth]{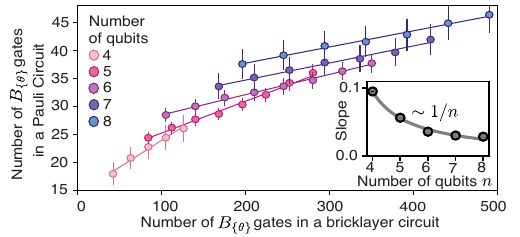}
  \caption{Expressiblity of Pauli rotations and Bricklayer circuits for $k=3$. These results are approximately equal to expressibility figure in the main text,
  but with larger error bars.
  }
  \label{fig:fig_circ_design_express_k3} 
\end{figure}

\vspace{5mm}

We also repeated our analysis for $k=3$, with results shown in Fig.~\ref{fig:fig_circ_design_express_k3}. These results are approximately equal to the $k=2$ case, but with larger error bars from sampling uncertainties [Fig.~\ref{fig:fig_circ_design_express_k3}]. In general, as $k$ increases, the error bars are expected to grow exponentially.

\section{Fault-tolerant Blind Quantum Algorithms}
\label{supp_sec:FTBQC}
In this section, we describe a method for implementing logical algorithms in a blind manner using a matter-based quantum computer. Our approach relies on constructing logical building blocks for blind computation, ensuring that the server remains oblivious to the quantum states being manipulated.

\subsection{Main Assumptions}

\begin{itemize}
    \item \textbf{Server's Role:} The server is responsible for generating logical states and conducting stabilizer checks. Stabilizer checks are performed by entangling and measuring auxiliary qubits locally. The measurement outcomes are sent to the client, who performs the full decoding of the logical state.
    \item \textbf{Server's Capabilities:} The server has fault-tolerant quantum computer (FTQC) with sufficient computational qubits to perform logical operations and communication qubits equipped with a strong optical interface for transmitting qubits.
    \item \textbf{Client's Role:} The client receives qubits from the server and performs measurements. It also decodes the logical algorithm using a classical computer and applies fast feedforward adjustments of measurement phases based on past measurement results.
\end{itemize}

\subsection{Blind Logical Blocks}

We propose constructing blind logical building blocks to execute the desired operations:

\begin{itemize}
    \item \textbf{1QC Block}: Implements logical single-qubit transversal gates.
    \item \textbf{2QC Block}: Implements logical entangling gates.
    \item \textbf{GT Block}: Implements logical gate teleportation.
\end{itemize}

\subsection{Quantum Error Correction and Stabilizer Checks}

Consider the surface code, which employs two types of measurement qubits: $X$-type and $Z$-type. After multiple layers of stabilizer checks, a 3D lattice structure forms (with time as the third dimension).
The client performs the decoding of this structure. For illustration, we use the surface code as a \([[n, k, d]]\) quantum error correction code (QECC), where \(k = 1\), \(n = 2d^2 - 1\), with \(d^2\) data qubits and \(d^2 - 1\) measurement qubits. 
Here, the $X$-type measurement qubits are responsible for checking $XXXX$ stabilizers, and $Z$-type measurement qubits for $ZZZZ$ stabilizers.

\subsection{Error modeling}
\begin{figure}[t]
  \centering
  \includegraphics[width=0.8\columnwidth]{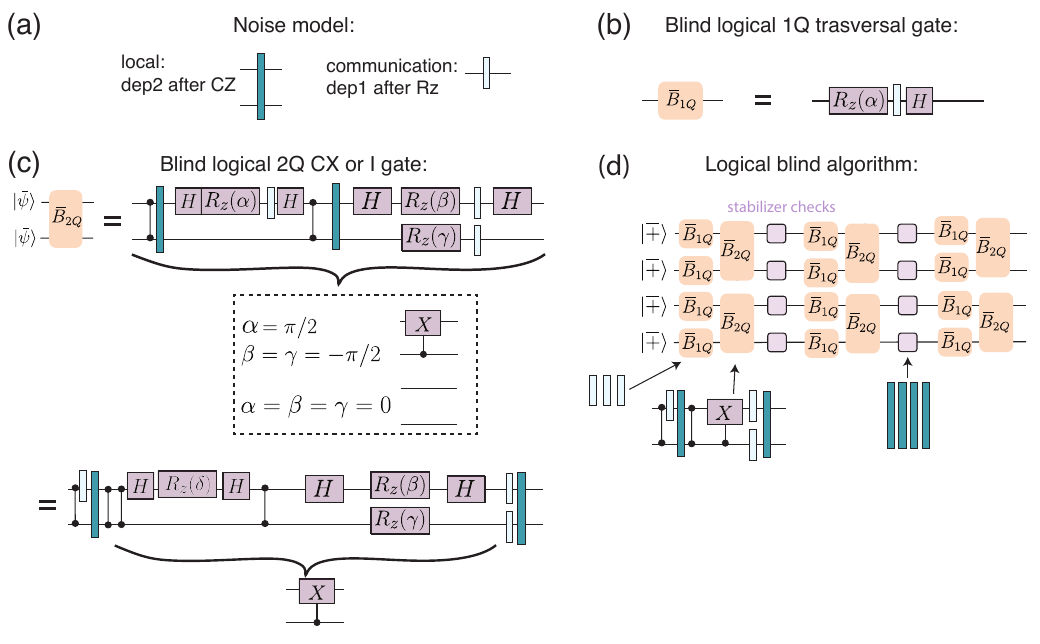}
  \caption{Error modeling in our blind logical algorithms simulations. For each gate type, we present the exact locations of the single qubit communication error and two-qubit local depolarizing error used in our simulations.
  }
  \label{SI_fig:QEC_simulations_error_model} 
\end{figure}

In our logical algorithms simulations, we simulate multiple logical surface codes, interacting via a random multi-layer circuit with a syndrome extraction (SE) layer after every gate layer.
Each local gate is followed by a 2-qubit depolarizing channel with probability $\epsilon_{\text{local}}$, and each delegated blind gate is followed by a single-qubit depolarizing channel with probability $\epsilon_{\text{comm}}$.
We illustrate our error modeling in Fig.~\ref{SI_fig:QEC_simulations_error_model}. 

\subsection{Logical Error Characterization}
\label{app:logical_error_per_gate}

To quantify the impact of noise on logical circuits, we define the algorithmic logical error rate, which ranges from \( 0 \) to \( p_{L,\text{max}} = 1 - \frac{1}{2^{N_{\text{q}}}} \), where \( p_{L,\text{max}} \) corresponds to the logical error rate of a maximally mixed logical state with \( N_{\text{q}} \) logical qubits.
We take a logical algorithm with $N_{\text{layers}}$ layers of gates, as illustrated in the main text. After each $1/N_{\text{rounds}}$ layer, there is a syndrome extraction (SE) round. For $N_{\text{rounds}} \geq 1$, there are $N_{\text{rounds}}$ SE rounds after each gate layer. In the gate layers, there could be blind gates, implemented using photons, or non-blind gates, implemented locally on the server. The fraction of blind gates is $R_{\text{h}}$, meaning that for $R=0$ all gates are local, and for $R=1$ all gates are blind. Therefore, in a layer with $N_{\text{gpl}}$ gates, there are $R\cdot N_{\text{gpl}}$ blind gates.

\textit{Logical Error Per Round and Per Gate:}
We define the logical error rate per round as:
\begin{equation}
\label{eq:logical_error_round_circuit}
    \frac{p_{L,\text{round}}}{p_{L,\text{max}}} = \left(1 - \left(1 - \frac{p_{L,\text{circuit}}}{p_{L,\text{max}}} \right)^{\frac{1}{N_{\text{layers}} \cdot N_{\text{rounds}}}} \right) \approx \frac{p_{L,\text{circuit}}}{N_{\text{layers}} \cdot N_{\text{rounds}}},
\end{equation}
where the approximation holds for \( p_{L,\text{circuit}} \ll p_{L,\text{max}} \). Here, \( N_{\text{rounds}} = 1 \) represents the number of syndrome extraction rounds per layer, and \( N_{\text{layers}} \) is the total number of gate layers in the logical circuit.

Similarly, the logical error per gate is defined as:
\begin{equation}
\label{eq:gate_round_relation}
    p_{L,\text{gate}} = 1 - (1 - p_{L,\text{round}})^{\frac{1}{N_{\text{gpl}}}},
\end{equation}
where \( N_{\text{gpl}} = 1.5 \cdot N_{\text{qubits}} \) accounts for the expected number of gates per layer in the random circuit.

Given communication errors only, we can write an expression for the blind gate error rates as a function of the logical round error rates. Here we can add the hiding ratio $R_{\text{h}}$:
\begin{equation}
\label{eq:gate_round_relation_ratio_hiding}
    p_{L,\text{blind gate}} = 1 - (1 - p_{L,\text{round}})^{\frac{1}{N_{\text{gpl}} \cdot R_{\text{h}}}},
\end{equation}
which will be useful to calculate the number of blind gates for circuits with both blind and non-blind gates.

Reversing this relationship, we obtain:
\begin{equation}
\label{eq:gate_round_inverse}
    p_{L,\text{round}} = 1 - (1 - p_{L,\text{gate}})^{N_{\text{gpl}}}.
\end{equation}

\begin{figure}[h]
\centering
\includegraphics[width=0.9\textwidth]{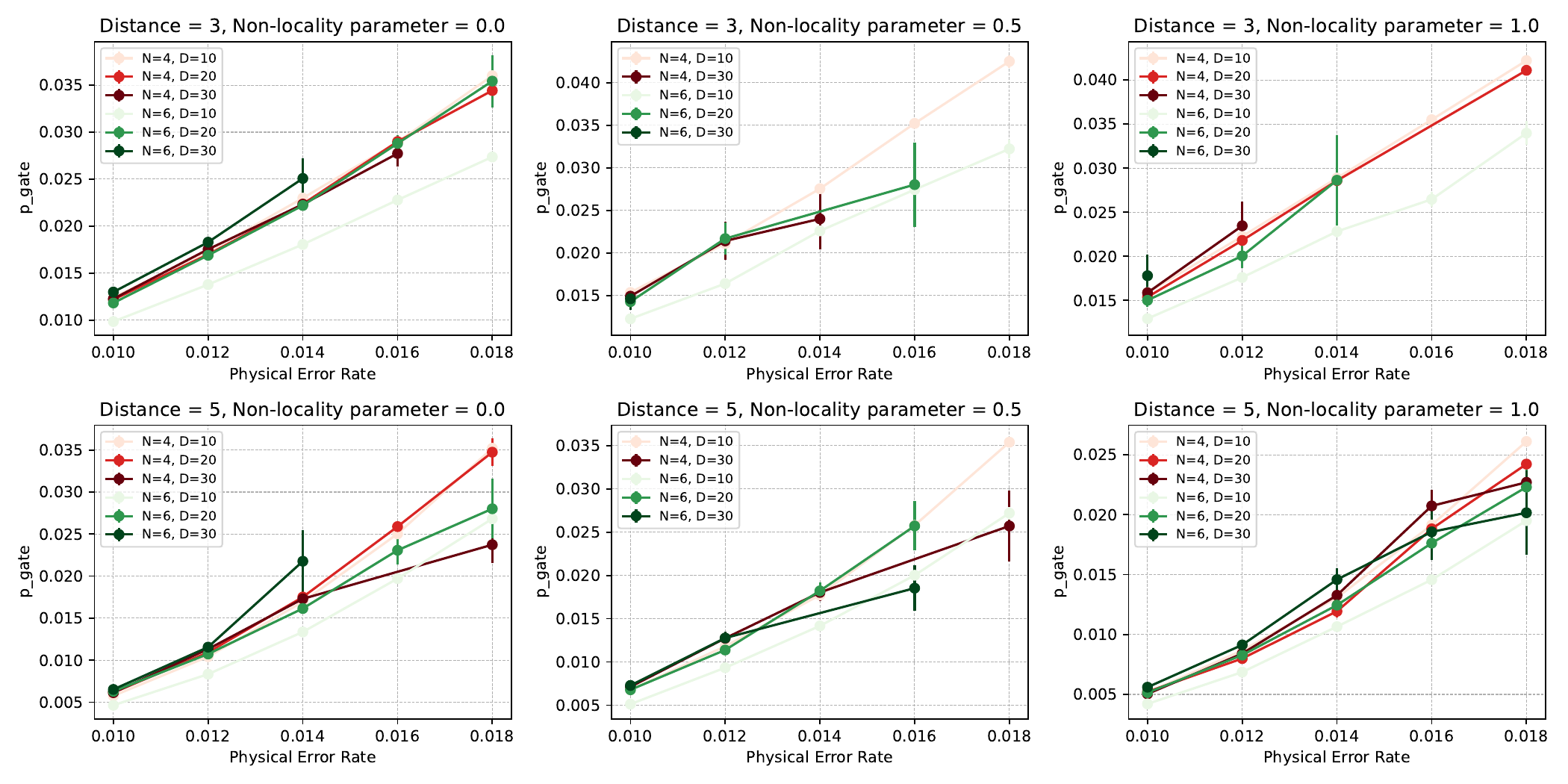}
\caption{Numerical validation of Eqs.~\eqref{eq:logical_error_round_circuit} and \eqref{eq:gate_round_relation}. We simulate a deep random logical circuit while varying code distances ($d=3,5$) and the non-locality fraction ($\Lambda=0,0.5,1$). For each parameter combination, we verify that varying the number of qubits ($N_{\text{qubits}}=4,6$) and the number of gate layers ($N_{\text{layers}}=10,20,30$) yields consistent scaling between $p_{\text{gate}}$ and the physical error rate. This confirms that, given a logical error per gate, the final fidelity of a deep random logical algorithm can be accurately estimated.}
\label{fig:QEC_numerical_proof}
\end{figure}

To validate these equations numerically, we compute \( p_{L,\text{gate}} \) as a function of the physical error rate for distances \( d = 3, 5 \), examining various parameter sets with \( N_{\text{layers}} = 10, 20, 30 \) and \( N_{\text{qubits}} = 4, 6 \). The results are presented in Fig.~\ref{fig:QEC_numerical_proof}.
The non-local fraction $\Lambda$ used in the figure is defined as:
\begin{equation}
    \Lambda = \frac{\epsilon_{\text{comm}}}{\epsilon_{\text{comm}} + \epsilon_{\text{local}}}.
\end{equation}
Meaning, for $\Lambda=0$, we consider local errors only; for $\Lambda=1$, we consider communication errors only; and for $\Lambda=0.5$, we consider both error types with the same probability.

\textit{Final Fidelity and Total Logical Gates:} 
We now express the total number of logical gates executed as a function of the final circuit fidelity. The total number of logical gates is given by:
\begin{equation}
    N_{\text{total gates}} = N_{\text{gpl}} \cdot N_{\text{layers}} \cdot N_{\text{rounds}}.
\end{equation}

Moreover, the total number of blind gates in the circuit is:
\begin{equation}
    N_{\text{total blind gates}} = R_{\text{h}} \cdot N_{\text{gpl}} \cdot N_{\text{layers}} \cdot N_{\text{rounds}},
\end{equation}
with $R_{\text{h}}$ the hiding fraction.

The final fidelity of the logical circuit can be written as:
\begin{equation}
\label{eq:fidelity_rounds_relation}
    F_{\text{L}} = 1 - p_{L,\text{circuit}} = 1 - p_{L,\text{max}} \left(1 - \left(1 - \frac{p_{L,\text{round}}}{p_{L,\text{max}}} \right)^{N_{\text{layers}} \cdot N_{\text{rounds}}} \right).
\end{equation}

For small \( p_{L,\text{round}} \), this simplifies to:
\begin{equation}
\label{eq:fidelity_approx}
    F_{\text{L}} \approx 1 - N_{\text{layers}} \cdot N_{\text{rounds}} \cdot p_{L,\text{round}} \approx (1 - p_{L,\text{round}})^{N_{\text{layers}} \cdot N_{\text{rounds}}}.
\end{equation}

Using Eq.~\eqref{eq:gate_round_relation}, we express the final fidelity as a function of a single logical blind gate:
\begin{equation}
\label{eq:fidelity_as_gates}
    F_{\text{L}} = 1 - p_{L,\text{max}} \left(1 - \left(1 - \frac{1 - (1 - p_{L,\text{gate}})^{N_{\text{gpl}}}}{p_{L,\text{max}}} \right)^{N_{\text{layers}} \cdot N_{\text{rounds}}} \right).
\end{equation}

Approximating for small \( p_{L,\text{gate}} \), we obtain:
\begin{equation}
    F_{\text{L}} \approx (1 - p_{L,\text{gate}})^{N_{\text{layers}} \cdot N_{\text{rounds}} \cdot N_{\text{gpl}}} = (1 - p_{L,\text{gate}})^{N_{\text{total gates}}}.
\end{equation}

Rearranging to express the total number of logical gates as a function of the final fidelity:
\begin{equation}
    \frac{N_{\text{total gates}}}{N_{\text{layers}} \cdot N_{\text{rounds}}} = \frac{\log{ \left(1 - p_{L,\text{max}} \left(1 - (1 - \frac{1 - F_{\text{L}}}{p_{L,\text{max}}})^{\frac{1}{{N_{\text{layers}} \cdot N_{\text{rounds}}}}} \right) \right)}}{\log{(1 - p_{L,\text{gate}})}}.
\end{equation}

For practical use, we further approximate:
\begin{equation}
    N_{\text{total gates}} \approx \frac{\log{F_{\text{L}}}}{\log{(1 - p_{L,\text{gate}})}}.
\end{equation}

This relation allows us to determine the maximum number of logical blind gates that can be executed while maintaining a final fidelity above a given threshold, e.g., \( F_{\text{final}} = 0.5 \). 
Given a hiding ratio $R_{\text{h}}<1$, we can utilize Eq.~\ref{eq:gate_round_relation_ratio_hiding} to calculate the total number of blind gates.

\subsection{Impact of Dark Counts and Photon Efficiency on Error Rates}

Photon measurement events in our quantum network protocol introduce errors through two primary mechanisms: dark counts and finite channel efficiency. 

Dark counts correspond to events in which no actual photon is present, yet the detector falsely registers a photon. Physically, a dark count event can be modeled as the environment effectively performing an unintended projective measurement on the photon's presence, causing the qubit to experience a random unknown rotation around the Z-axis, $R_z(\theta)$, with $\theta$ uniformly distributed. Such a scenario can be described by a fully Z-dephasing channel $\mathcal{D}_Z(\rho)$, given by:
\begin{equation}
    \mathcal{D}_Z(\rho) = \frac{1}{2}(\rho + Z \rho Z).
\end{equation}
This fully dephasing channel eliminates off-diagonal elements of the density matrix in the computational (Z) basis, thereby destroying coherence in the X-Y plane. Consequently, any intended Z-basis rotation gate $R_z(\phi)$ preceding this event becomes irrelevant, as the channel annihilates the coherence it would otherwise introduce. Formally, $\mathcal{D}_Z \circ R_z(\phi) = \mathcal{D}_Z$, meaning the previous rotation does not alter the outcome of the noise process.

Real photon detection events, however, introduce errors through a depolarizing channel, parameterized by the communication error rate $\epsilon_\text{comm}$. The complete noise channel upon detecting a photon (real or dark count) can thus be represented as a mixture:
\begin{equation}
    \mathcal{E}(\rho) = \frac{p_\text{dark}}{p_\text{dark} + \eta}\mathcal{D}_Z(\rho) + \frac{\eta}{p_\text{dark} + \eta}\left[(1 - \epsilon_\text{comm})\rho + \epsilon_\text{comm}\mathcal{D}_\text{dep}(\rho)\right],
\end{equation}
where $\eta$ is the photon efficiency and $\mathcal{D}_\text{dep}$ is the depolarizing channel, which introduces an equal probability of X, Y, and Z errors.

The effective error probability per photon detection event is thus:
\begin{equation}
    p_\text{err} = \frac{p_\text{dark}}{p_\text{dark} + \eta} + \epsilon_\text{comm}\frac{\eta}{p_\text{dark} + \eta}.
\end{equation}

For negligible communication errors ($\epsilon_\text{comm}\rightarrow 0$), the dark count error dominates, imposing a fundamental upper bound on the operational distance of the protocol via the condition:
\begin{equation}
    p_\text{thresh} > \frac{p_\text{dark}}{p_\text{dark} + \eta}.
\end{equation}

With typical state-of-the-art dark count rates $p_\text{dark} \sim 2\times10^{-7}$ and photon efficiency $\eta= 0.885~10^{-L/50}$ (for a cavity based setting), the maximum achievable distance is approximately $L \approx 280$~km.

It is important to highlight that threshold error rates for Z-dephasing channels are typically significantly higher (around 2–3 times larger) than those for depolarizing channels, due to the reduced complexity and severity of errors introduced. Therefore, our threshold estimation based on the depolarizing noise alone serves as a conservative benchmark. Our methodology inherently captures the additional error contribution from dark counts, ensuring robustness even under realistic operational conditions.

Practically, at these large distances, the probability of photon detection events decreases dramatically, requiring around $10^7$ attempts per successful detection. At such regimes, other practical considerations and additional decoherence processes become dominant.

Finally, inter-node photon loss probabilities ($2.5\times 10^{-4}$ to $2.5\times10^{-2}$~\cite{wei:2024}) are comfortably below our computed thresholds, confirming the sufficiency of our error correction strategy under realistic experimental conditions. Should dark counts or related issues escalate in significance, our model remains adaptable to more detailed analyses.

\bibliography{references2}